\documentclass[twocolumn]{aastex63}
\usepackage{graphicx}
\usepackage{hyperref}
\usepackage{amsmath}

\received{XXX}
\revised{YYY}
\accepted{ZZZ}

\submitjournal{ApJ Letters}

\newcommand{\us}{$\mu$s}
\newcommand{\nicer}{{\it NICER}}
\newcommand{\chimepsr}{CHIME/Pulsar}
\newcommand{\pbdot}{$\dot{P}_{\rm b}$}
\newcommand{\pbdottotal}{($\dot{P}_{\rm b})_{\rm obs}$ = $(\dot{P}_{\rm b})_{\rm GR}$ + $(\dot{P}_{\rm b})_{\rm DR}$ + $(\dot{P}_{\rm b})_z$ + $(\dot{P}_{\rm b})_\mu$}
\newcommand{\tempo}{{\tt tempo}}
\newcommand{\nanopipe}{{\tt nanopipe}}
\newcommand{\psrchive}{{\tt psrchive}}
\newcommand{\pp}{{\tt PulsePortraiture}}

\begin{document}
\title{Refined Mass and Geometric Measurements of the High-Mass PSR J0740+6620}

\shorttitle{Extended Observations of PSR J0740+6620}

%% load in authors.
\author[0000-0001-8384-5049]{E. Fonseca}
    \affiliation{Department of Physics, McGill University, 3600 rue University, Montr\'eal, QC H3A 2T8, Canada}
    \affiliation{McGill Space Institute, McGill University, 3550 rue University, Montr\'eal, QC H3A 2A7, Canada}
    \affiliation{Department of Physics and Astronomy, West Virginia University, Morgantown, WV 26506-6315, USA}
    \affiliation{Center for Gravitational Waves and Cosmology, Chestnut Ridge Research Building, Morgantown, WV 26505, USA}
\author[0000-0002-6039-692X]{H.~T. Cromartie}
    \affiliation{Cornell Center for Astrophysics and Planetary Science and Department of Astronomy, Cornell University, Ithaca, NY 14853, USA}
    \affiliation{NASA Hubble Fellowship Program Einstein Postdoctoral Fellow}
\author[0000-0001-5465-2889]{T.~T. Pennucci}
    \affiliation{National Radio Astronomy Observatory, 520 Edgemont Road, Charlottesville, VA 22903, USA}
    \affiliation{Institute of Physics, E\"otv\"os Lor\'an University, P\'azm\'any P. s. 1/A, 1117 Budapest, Hungary}
\author[0000-0002-5297-5278]{P.~S. Ray}
    \affiliation{U.S. Naval Research Laboratory, Washington, DC 20375, USA}
\author[0000-0002-8139-8414]{A.~Yu. Kirichenko}
    \affiliation{Universidad Nacional Aut\'{o}noma de M\'{e}xico, Instituto de Astronom\'{i}a, AP 106,  Ensenada 22800, BC, M\'{e}xico}
    \affiliation{Ioffe Institute, Politekhnicheskaya 26, St. Petersburg, 194021, Russia}
\author[0000-0001-5799-9714]{S.~M. Ransom}
    \affiliation{National Radio Astronomy Observatory, 520 Edgemont Road, Charlottesville, VA 22903, USA}
\author{P.~B. Demorest}
    \affiliation{National Radio Astronomy Observatory, P.O. Box O, Socorro, NM 87801, USA}
\author[0000-0001-9784-8670]{I.~H. Stairs}
    \affiliation{Department of Physics \& Astronomy, University of British Columbia, 6224 Agricultural Road, Vancouver, BC V6T 1Z1, Canada}
\author{Z. Arzoumanian}
    \affiliation{Astrophysics Science Division, Code 662, NASA Goddard Space Flight Center, Greenbelt, MD 20771, USA}
\author[0000-0002-9049-8716]{L. Guillemot}
    \affiliation{Laboratoire de Physique et Chimie de l'Environnement et de l'Espace -- Universit\'e d'Orl\'eans / CNRS, F-45071 Orl\'eans Cedex 02, France}
    \affiliation{Station de radioastronomie de Nan\c{c}ay, Observatoire de Paris, CNRS/INSU, F-18330 Nan\c{c}ay, France}
\author[0000-0002-4140-5616]{A. Parthasarathy}
    \affiliation{Max Planck Institute for Radio Astronomy, Auf dem H \ "{u} gel 69, D-53121 Bonn, Germany}
\author[0000-0002-0893-4073]{M. Kerr}
    \affiliation{U.S. Naval Research Laboratory, Washington, DC 20375, USA}
\author{I. Cognard}
    \affiliation{Laboratoire de Physique et Chimie de l'Environnement et de l'Espace -- Universit\'e d'Orl\'eans / CNRS, F-45071 Orl\'eans Cedex 02, France}
    \affiliation{Station de radioastronomie de Nan\c{c}ay, Observatoire de Paris, CNRS/INSU, F-18330 Nan\c{c}ay, France}
\author[0000-0003-2745-753X]{P.~T. Baker}
    \affiliation{Department of Physics and Astronomy, Widener University, Chester, PA 19013, USA}
\author[0000-0003-4046-884X]{H. Blumer}
    \affiliation{Department of Physics and Astronomy, West Virginia University, Morgantown, WV 26506-6315, USA}
    \affiliation{Center for Gravitational Waves and Cosmology, Chestnut Ridge Research Building, Morgantown, WV 26505, USA}
\author[0000-0003-3053-6538]{P.~R. Brook}
    \affiliation{Department of Physics and Astronomy, West Virginia University, Morgantown, WV 26506-6315, USA}
    \affiliation{Center for Gravitational Waves and Cosmology, Chestnut Ridge Research Building, Morgantown, WV 26505, USA}
\author[0000-0002-2185-1790]{M. DeCesar}
    \affiliation{Department of Physics and Astronomy, George Mason University, Fairfax, VA 22030, resident at U.S. Naval Research Laboratory, Washington, D.C. 20375, USA}
\author[0000-0001-8885-6388]{T. Dolch}
    \affiliation{Department of Physics, Hillsdale College, 33 E. College Street, Hillsdale, MI 49242, USA}
    \affiliation{Eureka Scientific, Inc. 2452 Delmer Street, Suite 100, Oakland, CA 94602-3017, USA}
\author[0000-0003-4098-5222]{F.~A. Dong}
    \affiliation{Department of Physics \& Astronomy, University of British Columbia, 6224 Agricultural Road, Vancouver, BC V6T 1Z1, Canada}
\author[0000-0001-7828-7708]{E.~C. Ferrara}
    \affiliation{Department of Astronomy, University of Maryland, College Park, MD 20742, USA}
    \affiliation{Center for Exploration and Space Studies (CRESST)}
    \affiliation{NASA Goddard Space Flight Center, Greenbelt, MD 20771, USA}
\author[0000-0001-5645-5336]{W. Fiore}
    \affiliation{Department of Physics and Astronomy, West Virginia University, Morgantown, WV 26506-6315, USA}
    \affiliation{Center for Gravitational Waves and Cosmology, Chestnut Ridge Research Building, Morgantown, WV 26505, USA}
\author{N. Garver-Daniels}
    \affiliation{Department of Physics and Astronomy, West Virginia University, Morgantown, WV 26506-6315, USA}
    \affiliation{Center for Gravitational Waves and Cosmology, Chestnut Ridge Research Building, Morgantown, WV 26505, USA}
\author[0000-0003-1884-348X]{D.~C. Good}
    \affiliation{Department of Physics \& Astronomy, University of British Columbia, 6224 Agricultural Road, Vancouver, BC V6T 1Z1, Canada}
\author[0000-0003-1082-2342]{R. Jennings}
    \affiliation{Department of Astronomy, Cornell University, Ithaca, NY 14853, USA}
\author[0000-0001-6607-3710]{M.~L. Jones}
    \affiliation{Center for Gravitation, Cosmology, and Astrophysics, Department of Physics, University of Wisconsin-Milwaukee, P.O. Box 413, Milwaukee, WI 53201, USA}
\author[0000-0001-9345-0307]{V.~M. Kaspi}
    \affiliation{Department of Physics, McGill University, 3600 rue University, Montr\'eal, QC H3A 2T8, Canada}
    \affiliation{McGill Space Institute, McGill University, 3550 rue University, Montr\'eal, QC H3A 2A7, Canada}
\author[0000-0003-0721-651X]{M.~T. Lam}
    \affiliation{School of Physics and Astronomy, Rochester Institute of Technology, Rochester, NY 14623, USA}
    \affiliation{Laboratory for Multiwavelength Astrophysics, Rochester Institute of Technology, Rochester, NY 14623, USA}
\author[0000-0003-1301-966X]{D.~R. Lorimer}
    \affiliation{Department of Physics and Astronomy, West Virginia University, Morgantown, WV 26506-6315, USA}
    \affiliation{Center for Gravitational Waves and Cosmology, Chestnut Ridge Research Building, Morgantown, WV 26505, USA}
\author[0000-0001-5373-5914]{J. Luo}
    \affiliation{Canadian Institute for Theoretical Astrophysics, University of Toronto, 60 St. George Street, Toronto, ON M5S 3H8, Canada}
\author[0000-0001-5481-7559]{A. McEwen}
    \affiliation{Center for Gravitation, Cosmology, and Astrophysics, Department of Physics, University of Wisconsin-Milwaukee, P.O. Box 413, Milwaukee, WI 53201, USA}
\author[0000-0002-2885-8485]{J.~W. McKee}
    \affiliation{Canadian Institute for Theoretical Astrophysics, University of Toronto, 60 St. George Street, Toronto, ON M5S 3H8, Canada}
\author{M.~A. McLaughlin}
    \affiliation{Department of Physics and Astronomy, West Virginia University, Morgantown, WV 26506-6315, USA}
    \affiliation{Center for Gravitational Waves and Cosmology, Chestnut Ridge Research Building, Morgantown, WV 26505, USA}
\author[0000-0002-4642-1260]{N. McMann}
    \affiliation{Department of Physics and Astronomy, Vanderbilt University, 2301 Vanderbilt Place, Nashville, TN 37235, USA}
\author[0000-0001-8845-1225]{B.~W. Meyers}
    \affiliation{Department of Physics \& Astronomy, University of British Columbia, 6224 Agricultural Road, Vancouver, BC V6T 1Z1, Canada}
\author[0000-0002-9225-9428]{A. Naidu}
    \affiliation{University of Oxford, Sub-Department of Astrophysics, Denys Wilkinson Building, Keble Road, Oxford, OX1 3RH, United Kingdom}
\author[0000-0002-3616-5160]{C. Ng}
    \affiliation{Dunlap Institute for Astronomy \& Astrophysics, University of Toronto, 50 St.~George Street, Toronto, ON M5S 3H4, Canada}
\author[0000-0002-6709-2566]{D.~J. Nice}
    \affiliation{Department of Physics, Lafayette College, Easton, PA 18042, USA}
\author[0000-0002-8826-1285]{N. Pol}
    \affiliation{Department of Physics and Astronomy, Vanderbilt University, 2301 Vanderbilt Place, Nashville, TN 37235, USA}
\author{H.~A. Radovan}
    \affiliation{Department of Physics, University of Puerto Rico, Mayag\"uez, PR 00681, USA}
\author[0000-0002-7283-1124]{B. Shapiro-Albert}
    \affiliation{Department of Physics and Astronomy, West Virginia University, Morgantown, WV 26506-6315, USA}
    \affiliation{Center for Gravitational Waves and Cosmology, Chestnut Ridge Research Building, Morgantown, WV 26505, USA}
\author[0000-0001-7509-0117]{C.~M. Tan}
    \affiliation{Department of Physics, McGill University, 3600 rue University, Montr\'eal, QC H3A 2T8, Canada}
    \affiliation{McGill Space Institute, McGill University, 3550 rue University, Montr\'eal, QC H3A 2A7, Canada}
\author[0000-0003-2548-2926]{S.~P. Tendulkar}
    \affiliation{Department of Astronomy and Astrophysics, Tata Institute of Fundamental Research, Mumbai, 400005, India}
    \affiliation{National Centre for Radio Astrophysics, Post Bag 3, Ganeshkhind, Pune, 411007, India}
\author[0000-0002-1075-3837]{J.~K. Swiggum}
    \affiliation{Department of Physics, Lafayette College, Easton, PA 18042, USA}
\author[0000-0001-9678-0299]{H.~M. Wahl}
    \affiliation{Department of Physics and Astronomy, West Virginia University, Morgantown, WV 26506-6315, USA}
    \affiliation{Center for Gravitational Waves and Cosmology, Chestnut Ridge Research Building, Morgantown, WV 26505, USA}
\author[0000-0001-5105-4058]{W.~W. Zhu}
    \affiliation{National Astronomical Observatories, Chinese Academy of Sciences, Beijing 100101, China}
%\shortauthors{E. Fonseca et al.}

%--------------------------------------------------------------------
\begin{abstract}
We report results from continued timing observations of PSR J0740+6620, a high-mass, 2.8-ms radio pulsar in orbit with a likely ultra-cool white dwarf companion. Our data set consists of combined pulse arrival-time measurements made with the 100-m Green Bank Telescope and the Canadian Hydrogen Intensity Mapping Experiment telescope. We explore the significance of timing-based phenomena arising from general-relativistic dynamics and variations in pulse dispersion. When using various statistical methods, we find that combining $\sim 1.5$ years of additional, high-cadence timing data with previous measurements confirms and improves upon previous estimates of relativistic effects within the PSR J0740+6620 system, with the pulsar mass $m_{\rm p} = 2.08^{+0.07}_{-0.07}$ M$_\odot$ (68.3\% credibility) determined by the relativistic Shapiro time delay. For the first time, we measure secular variation in the orbital period and argue that this effect arises from apparent acceleration due to significant transverse motion. After incorporating contributions from Galactic differential rotation and off-plane acceleration in the Galactic potential, we obtain a model-dependent distance of $d = 1.14^{+0.17}_{-0.15}$ kpc (68.3\% credibility). This improved distance confirms the ultra-cool nature of the white dwarf companion determined from recent optical observations. We discuss the prospects for future observations with next-generation facilities, which will likely improve the precision on $m_{\rm p}$ for J0740+6620 by an order of magnitude within the next few years. 
\end{abstract}


%--------------------------------------------------------------------
\section{Introduction}
\label{sec:intro}

The masses and radii of neutron stars are a central focus in many observational high-energy experiments. Members of the neutron-star population with the largest masses are particularly important. A key application of these measurements is in testing proposed neutron-star equations of state (EoSs), of which there are many that incorporate baryonic and/or ``exotic" compositions not accessible through terrestrial experiments \citep{of16}. For many decades, the primary means of measuring precise neutron star masses was through timing of radio pulsars in orbital systems that exhibit relativistic dynamical processes \citep[e.g.,][]{sta03}. Observatories sensitive to X-ray and gravitational radiation have begun to yield similar estimates of masses and radii, propelling statistical inference of fundamental physics and EoS constraints from neutron stars into a new and exciting era \citep[e.g.,][]{aaa+18,mld+19,rwb+19}.

A recent example of important observational constraints arises from the timing of PSR J0740+6620, a 2.8-ms millisecond pulsar (MSP) in a near-circular, 4.7-day orbit discovered by the Green Bank North Celestial Cap Survey \citep[GBNCC;][]{lsk+18}. Measurement of the relativistic Shapiro time delay \citep{sha64} in the PSR J0740+6620 system allowed for an estimate of the pulsar mass, $m_{\rm p} = 2.14^{+0.10}_{-0.09}$ M$_\odot$ \citep[68.3\% credibility;][]{cfr+20}. An optical search for the white-dwarf companion to PSR J0740+6620 yielded a detection of a likely ultra-cool counterpart \citep{bkk+19}, though a firm association is dependent on a reliable measure of distance that was not obtainable at the time. These radio and optical-photometric data were shown to be consistent with binary-stellar-evolution simulations of a progenitor system undergoing conservative mass transfer within a Hubble time, regardless of whether irradiation feedback was prominent in system evolution \citep{enbd20}.

Along with PSRs J0348+0423 \citep{afp+13} and J1614$-$2230 \citep[e.g.,][]{dpr+10}, PSR J0740+6620 resides at the high-mass extremum of the known Galactic pulsar population. In addition, and while subject to greater modeling uncertainty, observations of ``redback" pulsars and their irradiated companions clearly suggest a substantial population of high-mass neutron stars \citep[e.g.,][]{vbk11,lsc18,rkf+21}. Gravitational-wave observations of GW190814 indicate a high-mass compact object that could be the heaviest neutron star known \citep{aaa+20}. A small yet growing population, high-mass neutron stars serve as the most important constraints on the maximum allowed masses and, by extension, the EoS models that predict these measurements \citep[e.g.,][]{hkw+20,rgr+20,trs20}. Further study of the high-mass pulsar population will also allow for an ensemble estimation of the maximum mass, as well as the study of distribution moments that reflect underlying formation processes, before or during the spin-up to millisecond rotational periods, that produce high-mass neutron stars \citep[e.g.,][]{ato+16}.

Recently, the Neutron Star Interior Composition Explorer (\nicer) measured the mass and radius of the isolated PSR J0030$+$0451 through modeling of X-ray thermal emission from hotspots on the neutron-star surface \citep{bgr+19,mld+19,rrw+19,rwb+19}. Constraints on neutron-star radii from \nicer{} observations are strengthened with {\it a priori} knowledge of their masses. For sufficiently bright pulsars in binary systems, typically with system inclination $i \gtrsim 60^\circ$, mass and geometric measurements can be achieved with high-precision pulsar timing. In late 2019, \nicer{} confirmed X-ray pulsations from PSR J0740+6620 \citep{wol+21} and began a dedicated observing program to infer the stellar compactness via modeling of its pulsed X-ray emission. We endeavor to improve upon the mass measurement of \citet{cfr+20} and constrain other physical parameters in support of \nicer{} observations and their constraints on the radius.

In this work, we present observations, analysis and results from ongoing timing observations of PSR J0740+6620 with two different telescopes. In Section \ref{sec:obs}, we overview the instruments and processing methods used to acquire pulse arrival-time data for PSR J0740+6620. In Section \ref{sec:method}, we describe the procedure used to simultaneously model arrival times obtained with different instruments and receivers. In Section \ref{sec:analysis}, we outline the theoretical framework used in order to obtain robust constraints on the Shapiro delay and distance to PSR J0740+6620. In Section \ref{sec:discussion}, we discuss various aspects of our analysis and results, including dispersion measure (DM) modeling and impacts our measurements have on recent optical observations. Finally, in Section \ref{sec:conclusions}, we summarize our results and speculate about future improvement of our measurements.

%--------------------------------------------------------------------
\section{Observations \& Reduction}
\label{sec:obs}
Continued observations of PSR J0740+6620 were conducted using the 100-m Green Bank Telescope (GBT) and the Canadian Hydrogen Intensity Mapping Experiment (CHIME) telescope. A summary of observing parameters and statistics is presented in Table \ref{tab:summary_raw_data}.

\begin{deluxetable*}{l|c|c|c}
    \tablehead{\colhead{Parameter} & \colhead{GBT/1400-MHz} & \colhead{GBT/820-MHz} & \colhead{CHIME}}
    \tablecaption{A summary of observing parameters and data sets of PSR J0740+6620 anaylzed in this work.\label{tab:summary_raw_data}}
    \startdata
    Time Range (MJD) & 56675--58945 & 56640--58944 & 58517--58975 \\
    Frequency Range (MHz) & 1151--1885 & 722--919 & 400--800 \\
    Number of Channels, raw & 512 & 128 & 1024 \\
    Number of Channels, downsampled & 64 & 64 & 32 \\
    Number of Observing Epochs & 187 & 142 & 263 \\
    Number of TOAs, narrowband & 8,208 & 5,191 & 4,862 \\
    Number of TOAs, wideband & 209 & 154 & 263 \\
    Min./Median/Max. $\sigma_{\rm TOA}$, narrowband ($\mu$s) & 0.24/2.65/32.7 & 0.16/3.16/24.9 & 0.49/5.21/62.8 \\
    Min./Median/Max. $\sigma_{\rm TOA}$, wideband ($\mu$s) & 0.05/0.39/2.04 & 0.04/0.51/3.10 & 0.14/0.97/3.28 \\
    \enddata
\end{deluxetable*}

\subsection{Telescopes}
Nearly all data from the single-dish GBT were acquired as part of the ongoing observing program of the North American Nanohertz Observatory for Gravitational Waves (NANOGrav\footnote{\href{http://nanograv.org}{nanograv.org}}), which monitors an array of nearly 100 MSPs for signatures of gravitational radiation at nanohertz frequencies. Details of the NANOGrav observing program are given by \citet{aab+21a,aab+21b}, and we provide a summary here.

At the GBT, raw telescope voltage data were coherently dedispersed and processed in real time through the Green Bank Ultimate Pulsar Processing Instrument \citep[GUPPI;][]{drd+08}. During its operation, GUPPI produced archives of time-integrated (or ``folded") pulse profiles, and for all elements of the Stokes polarization vector, over 10-s time bins. Folded profiles were collected using two different radio receivers at the GBT, with the receivers centred at radio frequencies of 820 MHz and 1400 MHz; these receivers possessed full bandwidths of 200 MHz and 800 MHz, respectively. Moreover, the spectrum of each folded profile was resolved over 128 and 512 channels when using the 820-MHz and 1400-MHz receivers, respectively. 

The GBT data set analyzed in this work also contains supplemental observations acquired by \citet{cfr+20} near and during superior conjunction in the PSR J0740+6620 binary system. Targeted observations have been shown to increase the measurement significance of the Shapiro time delay, which is at its maximum amplitude during superior conjunction \citep[orbital phase of 0.25, relative to the ascending node; e.g.,][]{dpr+10,pen15}. The targeted observations of PSR J0740+6620 occurred over three sessions: one 6-hour session occurred on MJD 58368 using the 820-MHz receiver during orbital phase of 0.25; and two 5-hour sessions were undertaken on MJDs 58448 and 58449 using the 1400-MHz receiver, corresponding to orbital phases of 0.15 and 0.25, respectively.

The CHIME telescope is a static radio interferometer comprised of four half-cylinder reflectors and a total of 1,024 dual-polarization antennas sensitive to the 400--800 MHz range. Digitized telescope voltage data from all CHIME feeds are coherently averaged using time-dependent phase delays corresponding to 10 different sources, instantaneously yielding 10 ``tracking" voltage timeseries. The CHIME/Pulsar backend \citep{abb+20} is a ten-node computing system that receives these 10 independent, complex voltage streams -- at resolutions of 2.56-$\mu$s and 1,024 frequency channels -- and uses {\tt dspsr} \citep{vb11} to perform real-time coherent dedispersion and folding. The resultant products are similar in structure to those acquired with GUPPI, though they contain 1,024 channels evaluated across the 400 MHz bandwidth of CHIME for each recorded profile. At the declination of PSR J0740+6620, acquisitions typically occur with $\sim 33$-minute duration in order to record a full transit at the CHIME telescope.

\subsection{Observing Cadences}
Under the nominal NANOGrav program, multi-frequency observations of PSR J0740+6620 with the GBT typically lasted for $\sim$25 minutes per receiver and occurred with a monthly cadence; observations with each receiver were acquired within $\sim$3 days of one another. After 2018 September 24 (i.e., MJD 58385), observations of PSR J0740+6620 were prioritized to occur during every scheduled NANOGrav session, which led to 3--5 multi-frequency observations within a one-week period of time during each month. See \citet{aab+21a,aab+21b} for additional discussion on the NANOGrav observing program.

Folded data were acquired with the \chimepsr{} coherent-dedispersion backend on a near-daily basis from 2019 February 3 to 2020 May 6 (i.e., MJD range 58517--58975). Exceptions to the near-daily cadence largely correspond to $\sim$one-week periods where the CHIME correlator ceased operation for software upgrades and hardware maintenance. Such activities only occurred $3-4$ times per year during the time span of CHIME data presented in this work.

\subsection{Offline Processing, Calibration and Downsampling}
All data acquired with the GBT and \chimepsr{} were processed offline using the \psrchive{}\footnote{http://psrchive.sourceforge.net/} suite of analysis utilities, through the use of separate \psrchive-based procedures tailored to the NANOGrav and \chimepsr{} data sets. For NANOGrav data on PSR J0740+6620, we used the \nanopipe{} pipeline\footnote{\url{https://github.com/demorest/nanopipe}} -- which is employed in ongoing NANOGrav analysis of all its sources -- for the following manipulations: mask-based excision of radio frequency interference (RFI); calibration of flux density and polarization information using on/off-source observations of the compact radio source J1445+0958 (B1442+101) modulated with a pulsed noise diode; downsampling to frequency resolution as low as 3.125 MHz per channel; and full integration of profiles across the duration of each scan. 

\chimepsr{} data were processed using a similar procedure, but with three modifications: (1) the Stokes profiles were not calibrated due to ongoing work in developing calibration methods for \chimepsr{} data; (2) spectra were downsampled to 32 channels with 12.5 MHz bandwidth, in order to retain adequate signal to noise (S/N) in as many channels as possible; and (3) $\sim\;1$ minute of profile data acquired before and after the transit of PSR J0740+6620 at CHIME were discarded prior to integration of profile data over time. The third modification was performed in order to minimize over-weighting of \chimepsr{} profiles recorded at low frequencies, where the primary beam of CHIME is wider and is thus sensitive to transiting pulsars for longer periods of time.

\subsection{Computation of Arrival Times}
\label{subsec:toas}

We generated two distinct sets of times-of-arrival (TOAs) and TOA uncertainties ($\sigma_{\rm TOA}$) from processed GBT and \chimepsr{} data: a set of frequency-resolved TOAs, which we refer to as ``narrowband'' TOAs; and a set of ``wideband'' TOAs, which extract a single arrival time measurement from each observation.

The narrowband TOAs were measured using the Fourier phase-gradient technique described in \citet{tay92}; the observed total-intensity (i.e., Stokes-$I$) profile from each frequency channel is cross-correlated with a de-noised template to measure a best-fit phase offset.  The template profile is constructed for each telescope and receiver combination by averaging all available per-receiver timing data, and then wavelet-smoothing the result \citep[e.g.,][]{abb+15}. These techniques of template and TOA generation form a standard protocol in pulsar astronomy, and were employed by \citet{cfr+20} in their analysis of the PSR J0740+6620 system. 

In this work, we obtained a maximum of 64 narrowband TOAs for each receiver bandwidth and all processed, RFI-cleaned observations. Narrowband TOAs derived from GBT data were further cleaned by removing outlier timing data using the NANOGrav analysis procedure outlined by \citet{aab+21a}, where TOAs were discarded from analysis based on: low profile fidelity (i.e., with a pulse profile S/N $<$ 8); corrupted calibration data; a probabilistic classification of outlier TOAs based on prior timing models \citep{vv17}; and manual TOA excision. Narrowband TOAs derived from \chimepsr{} data were cleaned using a similar procedure, though only consisted of rejecting low-S/N profiles and performing small amounts of manual TOA excision. Work is underway to fully integrate \chimepsr{} data into the NANOGrav analysis infrastructure and will be presented in future analyses.

\begin{figure}
    \centering
    \includegraphics[scale=0.062]{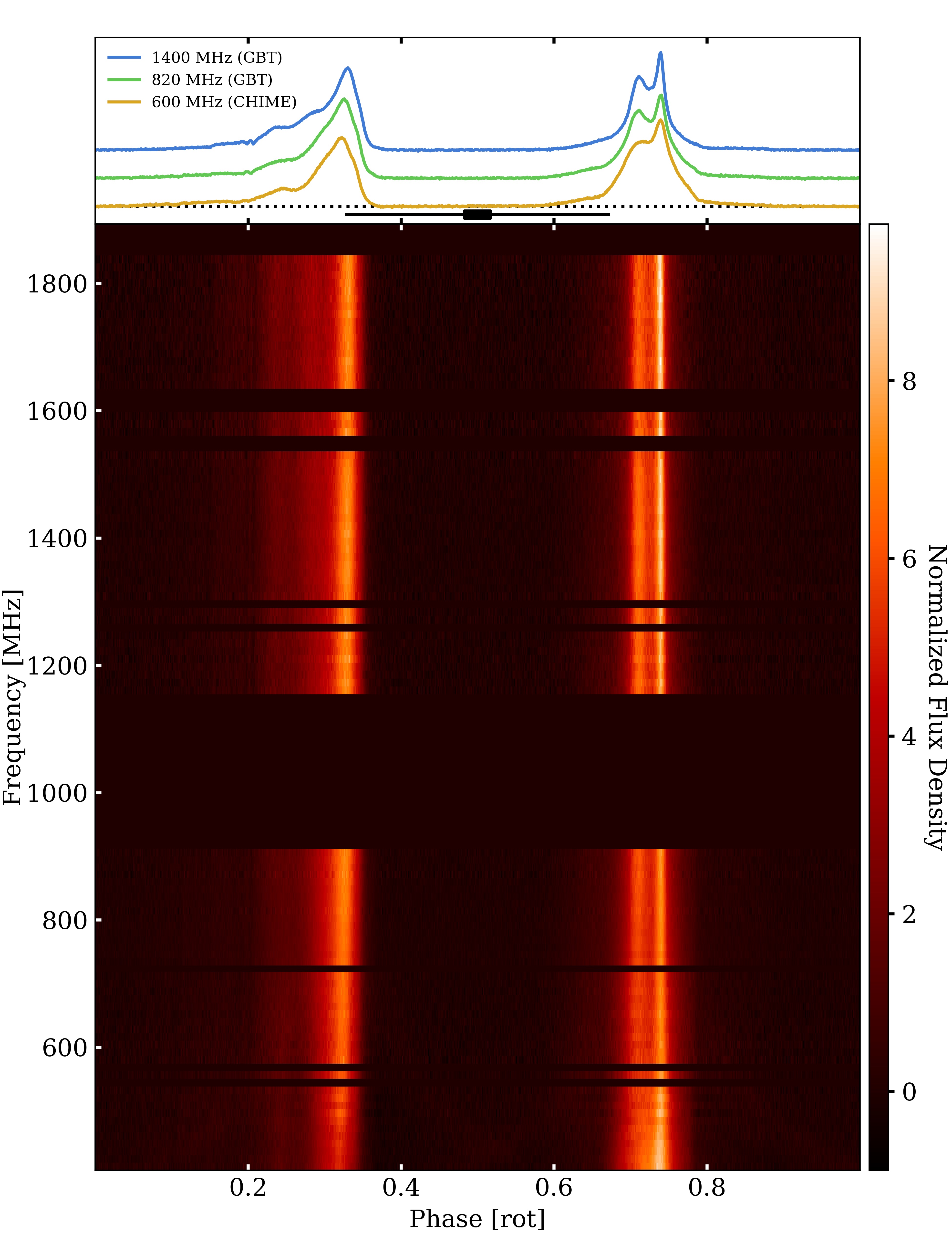}
    \caption{A composite spectrum of mean pulse profiles for PSR J0740+6620 across the three observed CHIME and GBT bands (Table~\ref{tab:summary_raw_data}). For each receiver, we integrated all data across observing time and down-sampled in frequency and phase to match the \chimepsr{} resolution (12.5\, MHz channels, 1024 phase bins).  The amplitudes of each profile are scaled such that all profiles have unit integrated flux density.  The profiles from each distinct band have been aligned for display against an arbitrary template of fixed Gaussian components, and the frequency-averaged profile from each band is displayed in the top panel. The horizontal line and bar in the top panel both denote timescale, with the line spanning 1 ms and the bar spanning 100 $\mu$s.  The large gap around 1000\, MHz is the gap between the GBT bands, and the narrow gap around 720\, MHz exists because the upper 20\% of the CHIME band was excised for this plot.  Other small gaps are places where there was persistent RFI or band roll-off.}
    \label{fig:0740_portrait}
\end{figure}

Wideband TOAs are a more recent innovation \citep{ldc+14,pdr14,pen19}, but have been incorporated in the recently released 12.5-yr NANOGrav data set \citep{aab+21b}. The wideband TOAs used for our analysis were generated using the procedure described by \citet{pdr14}\footnote{\url{https://github.com/pennucci/PulsePortraiture}}, which we briefly summarize as follows.  A single, wideband TOA and instantaneous DM were estimated for each observation using a generalized Fourier phase-gradient algorithm that constrains the frequency-dependent phase offsets. In order to derive a ``portrait" template for each receiver band, we first integrated all available data and computed a high-S/N template; the three mean portraits are conjoined and shown in Figure~\ref{fig:0740_portrait}. The de-noised portraits, computed for each receiver band from the averaged templates, were then derived through a combination of principal component analysis, wavelet smoothing, and spline interpolation. The only TOAs excised from the wideband data set were those from low-S/N observations \citep[with S/N $<$ 25; see][for further explanation]{aab+21b}.

%--------------------------------------------------------------------
\section{Timing Methods}
\label{sec:method}
% insert a couple of paragraphs here describing the timing formula, tempo, etc.
We constructed separate models of timing variations in PSR J0740+6620 using the narrowband and wideband data sets described in Section \ref{subsec:toas}. For each TOA set, we combined NANOGrav and \chimepsr{} TOAs and used the \tempo{} pulsar-timing package\footnote{\url{https://ascl.net/1509.002}} for conversion and generalized least-squares modeling \citep{chc+11} in terms of parameters that are specific to relevant physical processes as well as instrumental effects. A summary of the timing procedure and model is provided below.

\subsection{Pre-Modeling Conversion of TOAs}
For the \chimepsr{} data set, a small portion of the earliest TOAs were impacted by timing offsets that occurred upon restarts of the CHIME correlator. These offsets arose due to improper packaging of timing data required for downstream, real-time processing and determination of UTC timestamps. The relevant correlator software was amended to resolve this error, and all \chimepsr{} data acquired after MJD 58600 no longer contain achromatic timing offsets between observing cycles. We used configuration metadata recorded by the CHIME software infrastructure to determine the exact time steps introduced by two correlator offsets, and used those values as corrections to data acquired between MJDs 58517--58543.\footnote{The values of these offsets are noted in the ``-to" metadata flag in the TOA data made publicly available in this work.}

\subsection{Construction of the Timing Model}

The barycentric TOAs were then modeled against a superposition of analytic time delays describing various effects of physical and instrumental origin \citep[e.g.,][]{bh86}. Parameters of these time delays consist of terms describing: pulsar spin and its spin-down evolution; astrometric effects, which yield measures of position, proper motion, and apparent variations of position due to timing parallax ($\varpi$); piecewise-constant estimates of DM across the time span of the combined data set, referred to below as ``DMX"; and orbital motion.

We used the ELL1 binary timing model \citep{lcw+01} to model the near-circular Keplerian motion of PSR J0740+6620 in its orbit with the companion white dwarf. The five Keplerian elements in our model consist of: the orbital period ($P_{\rm b}$); the projected semi-major axis along the line of sight ($x$); the epoch of passage through the ascending node of the binary system ($T_{\rm asc}$); and the ``Laplace-Lagrange" eccentricity parameters, $\epsilon_1$ and $\epsilon_2$, that describe departures from circular motion. The ELL1 parameters are related to the traditional orbital elements -- eccentricity ($e$), argument of periastron ($\omega$), and epoch of passage through periastron ($T_0$) -- in the following manner:

\begin{align}
    e &= \sqrt{\epsilon_1^2 + \epsilon_2^2} \\
    \omega &= \arctan{\big(\epsilon_1/\epsilon_2\big)} \\
    T_0 &= T_{\rm asc} + P_{\rm b} \times \omega / (2\pi).
\end{align}

\noindent Following \citet{cfr+20}, we also included the Shapiro delay in our modeling of the extended data set, described by the ``range" ($r$) and ``shape" ($s$) parameters. While $s = \sin i$, where $i$ is the inclination of the binary system relative to the plane of the sky, the exact expression of $r$ depends on the assumed theory of gravitation \citep{dt92}; we assumed that general relativity is valid at the level of timing precision achieved in this work, which leads to $r = {\rm T}_\odot m_{\rm c}$, where T$_\odot = G{\rm M}_\odot/c^3$ = 4.925490947 $\mu$s and $m_{\rm c}$ is the mass of the companion star. With estimates of $m_{\rm c}$ and $\sin i$, the pulsar mass ($m_{\rm p}$) is determined by the Keplerian mass function,

\begin{equation}
    \label{eq:massfunc}
    f_m = \frac{4\pi^2}{{\rm T}_\odot}\frac{x^3}{P_{\rm b}^2} = \frac{(m_{\rm c}\sin i)^3}{(m_{\rm p} + m_{\rm c})^2}.
\end{equation}

The significance of all model parameters was determined using the {\it F}-test criterion, as described in \citet{aab+21a} for standard NANOGrav analysis, when comparing best-fit models that included or ignored each relevant set of parameters. We chose a $p$-value threshold of 0.0027 to reject the null hypothesis that adding or removing a parameter results in a variation of the residuals consistent with noise. Under Gaussian interpretation, a threshold of 0.0027 corresponds to a $3\sigma$ deviation.  Parameters were included in the model if their inclusion resulted in rejection of the null hypothesis based on this test. We explored fitting for secular variations in the orbital elements using the {\it F}-test criterion, and found that the time derivative of the orbital period (\pbdot) possessed sufficient statistical significance for inclusion as a degree of freedom. We interpret and discuss the implications of the \pbdot{} measurement in several sections below.

\subsection{Frequency Evolution of Pulse Profiles}
\label{subsec:methods_profile}

One of the key differences between the commonly-used narrowband methods of TOA estimation and wideband TOA estimation is their treatment of intrinsic profile evolution across receiver bandpasses.  A single, achromatic template profile is used to measure our narrowband TOAs.  Systematic time delays arise in the TOAs as a function of frequency in the presence of significant variation of the profile shape, which can be seen in Figure~\ref{fig:0740_portrait}.  We modeled these frequency-dependent (FD) variations in our narrowband TOA data set using the heuristic model developed by \citet{abb+15}, where the associated time delay $\Delta t_{\rm FD} = \sum_{i=1}^n c_i\ln{(f / 1 \mathrm{GHz})}^i$. The coefficients $c_i$ are free parameters in \tempo{}, and we included three FD coefficients -- with the number of coefficients determined using the {\it F}-test procedure described above -- in all timing models derived from narrowband TOAs.

For the wideband TOAs, the use of a high-fidelity frequency-dependent template ameliorates the timing biases introduced by profile evolution.  The FD parameters were not significant when checked by the F-test.  However, it is necessary to include three timing model parameters that quantify an offset in the average DM measured in each receiver band due to template profile misalignment; these DM parameters are analogous to the common phase ``JUMP'' parameters in \tempo{} \citep[see the ``DMJUMP'' parameters described in][]{aab+21b}.

\subsection{Analysis of Noise Properties}
\label{subsec:methods_noise}

A common procedure in pulsar timing is the analysis of noise properties in TOA residuals (the differences between measured TOAs and their values predicted by the best-fit timing model). This type of analysis involves using a number of heuristic parameters such that the uncertainty-normalized residuals are normally distributed with unit variance.  One set of these parameters adjusts the TOA uncertainties by a multiplicative factor ($F_k$), and a second set of these parameters is added to the TOA uncertainties in quadrature ($Q_k$). The subscript $k$ denotes the unique data subset for a set of frontend/backend combinations, and corresponds to the three subsets presented in Table \ref{tab:summary_raw_data} for this work. The factor $F_k$ accounts for incorrect estimation of $\sigma_{\rm TOA}$ due to mismatch between the data and template profiles, while $Q_k$ characterizes underlying additive white noise (i.e., statistical fluctuations with constant power spectral density across all times and frequencies).  A third set of parameters ($C_k$) encapsulates underlying noise processes that are uncorrelated in time but instead are fully correlated across observing frequency, and thus applies to simultaneously observed, multi-frequency TOAs.  One source for $C_k$ is the stochastic scatter introduced from pulse ``jitter" \citep[e.g.,][]{abb+15,lcc+17}.

We used the same methodology employed by \citet{aab+21a,aab+21b} for determining the optimal values of \{$F_k, Q_k, C_k$\} for the three narrowband data subsets listed in Table \ref{tab:summary_raw_data}. For all timing solutions, we used the {\tt enterprise} Bayesian pulsar timing suite \citep{evtb20} for sampling the  \{$F_k, Q_k, C_k$\} terms while marginalizing over all other free parameters in the timing model.  Following \citet{aab+21b}, a slightly different noise model is used in the wideband analysis. $C_k$ cannot be modeled for wideband TOAs because there are no simultaneously observed multi-frequency TOAs.  However, in addition to \{$F_k, Q_k$\} for the TOA uncertainties, analogous error-scaling values of $F_k$ for the DM uncertainties were also modeled. In all cases, the timing solution was refined until convergence using {\tt tempo} by applying the noise model from the {\tt enterprise} analysis.

We also explored the significance of temporal correlations in TOA residuals for PSR J0740+6620. Such ``red" noise has been seen in other NANOGrav MSPs and is understood to reflect irregularities in pulsar-spin rotation \citep{sc10}. However, we determined that red noise is not prominent in the timing of PSR J0740+6620, as its estimation with {\tt enterprise} yielded a Bayes significance factor of $\sim1$. We therefore did not include terms that quantify red noise in the noise model for PSR J0740+6620, as NANOGrav sets a Bayes factor threshold of 100 for including red-noise parameters into timing models \citep[e.g.,][]{abb+15}.

%---------------------------------------------
\section{Analysis} 
\label{sec:analysis}

\begin{deluxetable*}{l|cc|c}
    \tablehead{\colhead{Parameter\tablenotemark{b}} & \colhead{Narrowband TOAs} & \colhead{Wideband TOAs} & \colhead{Difference ($\sigma_{\rm max}$)\tablenotemark{c}}\label{tab:timing_model}}
    \tablecaption{A summary of timing parameters for PSR J0740+6620 when using DMX to estimate DM values\tablenotemark{a} in 6.5-day bins.}
    \startdata
    \multicolumn{4}{c}{Astrometry} \\
    \hline
    Ecliptic longitude, $\lambda$ (degrees) \dotfill & 103.759135333(12) & 103.759135338(13) & 0.4 \\
    Ecliptic latitude, $\beta$ (degrees) \dotfill & 44.102478368(13) & 44.102478361(13) & 0.5 \\
    Proper motion in $\lambda$, $\mu_\lambda$ (mas yr$^{-1}$) \dotfill & $-$2.735(14) & $-$2.737(15) & 0.1 \\
    Proper motion in $\beta$, $\mu_\beta$ (mas yr$^{-1}$) \dotfill & $-$32.48(2) & $-$32.48(2) & 0.0 \\
    Timing parallax, $\varpi$ (mas) \dotfill & 1.04(18) & 0.87(19) & 0.9 \\
    \hline
    \multicolumn{4}{c}{Spin} \\
    \hline
    Spin frequency, $\nu_{\rm s}$ (s$^{-1}$) \dotfill & 346.5319964608338(3) & 346.5319964608337(3) & 0.0 \\
    Time rate of change in frequency, $\dot{\nu}_{\rm s}$ ($10^{-15}$ s$^{-2}$) \dotfill &  $-$1.463874(11) & $-$1.463870(11) & 0.4 \\
    \hline
    \multicolumn{4}{c}{Binary Motion} \\
    \hline
    Orbital period, $P_{\rm b}$ (days) \dotfill & 4.76694461933(8) & 4.76694461936(8) & 0.4 \\
    Projected semi-major axis, $x$ (lt-s) \dotfill & 3.97755608(10) & 3.97755607(11) & 0.0 \\
    First Laplace-Lagrange eccentricity parameter, $\epsilon_1$ ($10^{-6}$) \dotfill & $-$5.68(3) & $-$5.70(3) & 0.7 \\
    Second Laplace-Lagrange eccentricity parameter, $\epsilon_2$ ($10^{-6}$) \dotfill & $-$1.833(18) & $-$1.840(19) & 0.3 \\
    Epoch of ascending-node passage, $T_{\rm asc}$ \dotfill & 57804.731308893(17) & 57804.731308895(18) & 0.1\\
    Time rate of change in period, $\dot{P}_{\rm b}$ (10$^{-12}$ s\,s$^{-1}$) \dotfill & 1.2(2) & 1.2(2) & 0.2 \\
    Companion mass, $m_{\rm c}$ (M$_\odot$) \dotfill & 0.251(5) & 0.253(6) & 0.3 \\
    Sine of inclination angle, $\sin i$ \dotfill & 0.99909(12) & 0.99908(13) & 0.1 \\
    \hline
    \multicolumn{4}{c}{Profile Evolution} \\
    \hline
    First coefficient of FD expansion, $c_1$ ($10^{-5}$) \dotfill & $-$3.2(2) & n/a & n/a \\
    Second coefficient of FD expansion, $c_2$ ($10^{-5}$) \dotfill & $-$2.0(3) & n/a & n/a \\
    Third coefficient of FD expansion, $c_3$ ($10^{-5}$) \dotfill & $-$1.0(2) & n/a & n/a \\
    \hline
    \multicolumn{4}{c}{Fit Configuration and Statistics} \\
    \hline
    Reference epoch for spin, astrometry (MJD) \dotfill & 57807 & 57807 & n/a \\ 
    Terrestrial clock standard \dotfill & TT(BIPM2019) & TT(BIPM2019) & n/a \\
    Solar system ephemeris \dotfill & DE438 & DE438 & n/a \\
    Barycentric time scale \dotfill & TDB & TDB & n/a \\
    Degrees of freedom \dotfill & 18,041 & 1,099 & n/a \\
    Goodness of fit, $\chi^2$ \dotfill & 18,183 & 1,072 & n/a \\
    Daily-averaged, weighted RMS ($\mu$s) \dotfill & 0.27 & 0.28 & n/a \\
    \enddata
    \tablenotetext{a}{We do not list DM values in this table due to the large number of DMX bins used to model DM variations. We refer the reader to Figures \ref{fig:residuals} and \ref{fig:residuals_zoom} for graphical representations of the best-fit DMX data.}
    \tablenotetext{b}{All parameter uncertainties, listed in parentheses, denote the 68.3\% (i.e., 1 $\sigma$) confidence intervals in the preceding digit(s) as obtained from \tempo{}.}
    \tablenotetext{c}{Differences between best-fit parameter estimates derived from narrowband and wideband TOAs are listed in units of the larger of the two statistical uncertainties ($\sigma_{\rm max}$).}
\end{deluxetable*}

\begin{figure*}[htb!]
    \centering
    \includegraphics[scale=1.0,width=2\columnwidth]{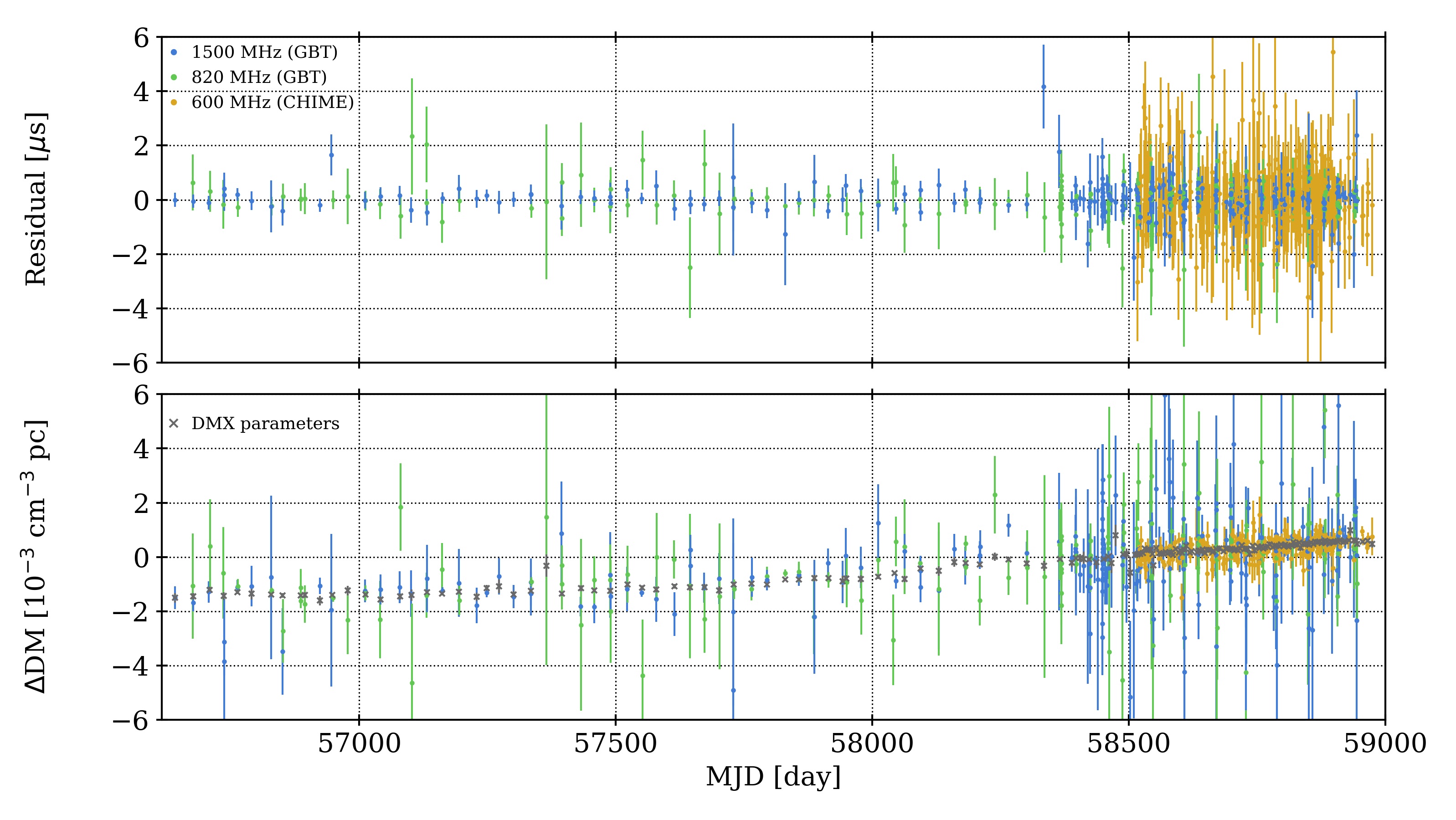}
    \caption{Timing residuals from the best-fit timing model for wideband TOAs (top), and DM time series (bottom).  The colors indicate the receiver and telescope of the observation in both panels.  The best-fit DMX model parameters (grey crosses) are determined by both the wideband DM measurements and the wideband TOAs.  The plotted $\Delta$DM values are offset from the nominal zero value of 14.9631 pc cm$^{-3}$. These wideband residual data and DMX model parameters correspond to the timing model that used the ``hybrid" DMX-binning scheme discussed in Section \ref{subsec:dmxmodels}.  A zoom-in of the \chimepsr{} data is provided in Figure~\ref{fig:residuals_zoom}.}
    \label{fig:residuals}
\end{figure*}

\begin{figure*}[htb!]
    \centering
    \includegraphics[scale=1.0,width=2\columnwidth]{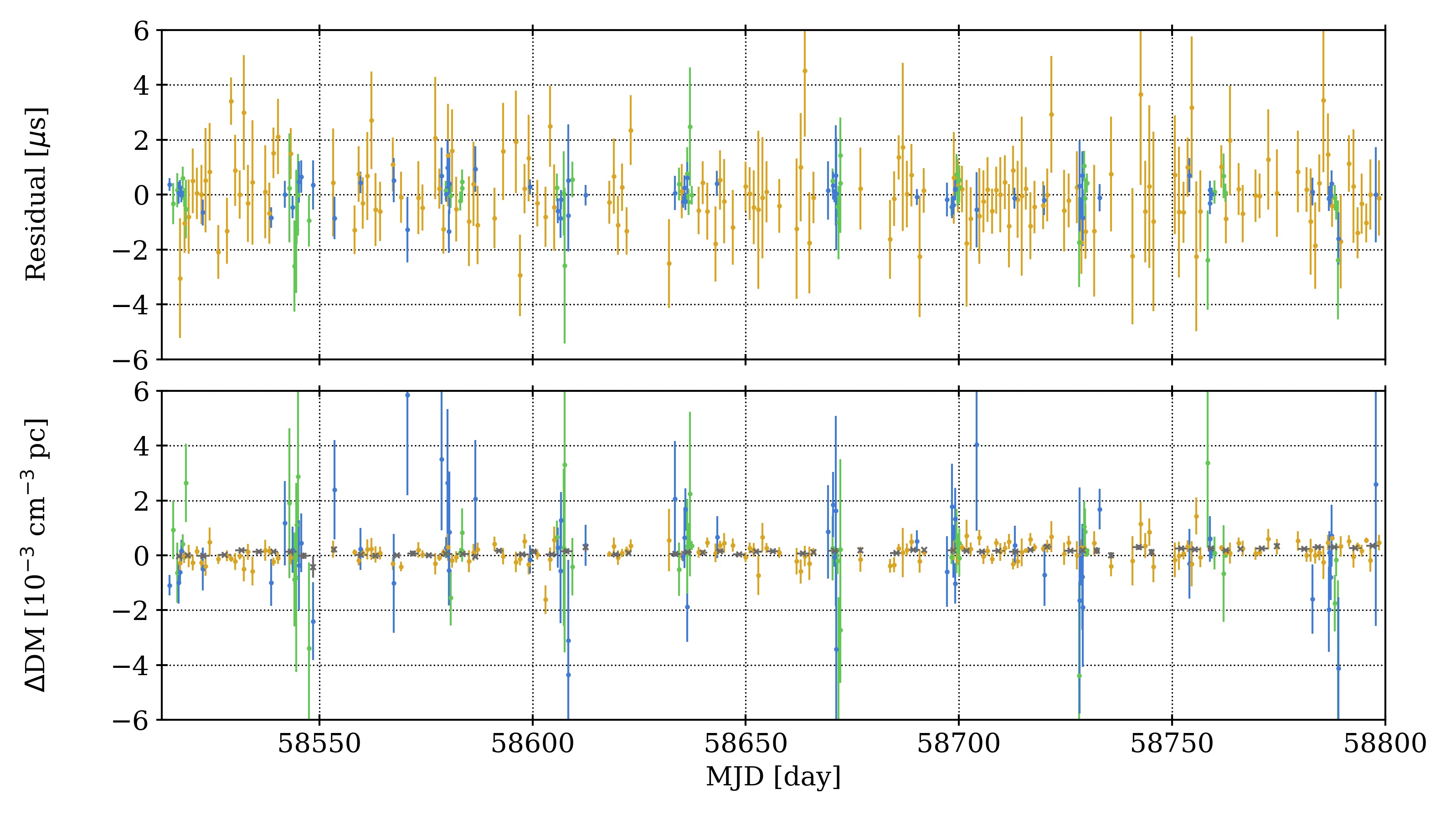}
    \caption{A zoomed-in view of the $\sim1$ yr overlap of GBT and \chimepsr{} data from Figure~\ref{fig:residuals}, sharing the same legend.  Worthy of note are: the relative cadences of observation; the smooth, shallow, and linear trend in DM over time; and the statistical weight of \chimepsr{} DM measurements as compared to its TOAs, relative to the GBT measurements.}
    \label{fig:residuals_zoom}
\end{figure*}

We explored several methods for statistical analysis described below to obtain robust credible intervals on key parameters of the PSR J0740+6620 system, using the modeling procedure described in Section \ref{sec:method}. The following analysis methods are agnostic to the use of narrowband or wideband TOAs. However, we chose to use the wideband TOAs in the work presented below given the consistency in parameters from narrowband and wideband TOAs in Table \ref{tab:timing_model}. Moreover, the following methods benefit from relaxed computational requirements afforded by the considerably smaller size ($\sim$1/20) of the wideband TOA data set.  The timing residuals and DM time series from the wideband data set are shown in Figure~\ref{fig:residuals}; a zoom-in of the overlap between the GBT and \chimepsr{} data sets is provided in Figure~\ref{fig:residuals_zoom}.

\subsection{DMX Models}
\label{subsec:dmxmodels}

Table \ref{tab:timing_model} presents two \tempo{} models, derived from independent fits to the narrowband and wideband TOA data sets, when using the same DMX method employed by \citet{cfr+20}. However, the different observing cadences within each TOA subset can lead to an uneven weighting of DMX estimates that impact the measurement of low-amplitude timing effects. In order to assess these impacts, we generated three sets of timing models that used slightly different DMX bins across the combined data set. Two of these DMX models used a single bin width of 3 and 6.5 days, respectively. In the third model, we used a hybrid scheme where the modeling of data acquired between MJDs 56640 and 58400 used a DMX model with bin size of 6.5 days, and all data acquired after MJD 58400 (when heightened-cadence observations began) were modeled with a DMX model using a bin size of 3 days.

These three DMX binning choices form a small subset of possible DM variation models. Alternative methods for modeling DM variation include the Fourier decomposition of frequency-resolved TOAs as implemented in the {\tt tempoNest} Bayesian analysis suite \citep{lah+14}. We nonetheless chose to use DMX as this method is able to resolve discrete, stochastic variations in DM that cannot otherwise be adequately modeled with, for example, a Taylor expansion of the DM. 

\subsection{Secular Variation in the Orbital Period}
\label{subsec:analysis_pbdot}
All fitted timing models exhibit a significant and consistent \pbdot. Our analysis represents the first time such variations have been detected in the PSR J0740+6620 system. Several mechanisms can yield apparent or intrinsic variations in $P_{\rm b}$, such as those due to: energy loss from quadrupole-order gravitational radiation \citep[(\pbdot)$_{\rm GR}$; e.g.,][]{dt92}; differential rotation in the Galaxy \citep[(\pbdot)$_\textrm{DR}$; e.g.,][]{nt95}; off-plane acceleration in the Galactic gravitational potential \citep[(\pbdot)$_z$;][]{kg89}; and apparent acceleration due to transverse motion \citep[(\pbdot)$_\mu$;][]{shk70}. We assumed that the total observed variation is therefore \pbdottotal, where 

\begin{align}
    \label{eq:pbdot_GR}
    (\dot{P}_{\rm b})_{\rm GR} & = -\frac{192\pi}{5}\frac{(n_{\rm b}{\rm T}_\odot)^{5/3}}{(1-e^2)^{7/2}}\frac{m_{\rm p}m_{\rm c}}{m_{\rm tot}^{1/3}}\bigg(1 + \frac{73}{24}e^2 + \frac{37}{96}e^4\bigg), \\
    \label{eq:pbdot_DR}
    (\dot{P}_{\rm b})_{\rm DR} &= -P_{\rm b}\cos b \bigg(\frac{\Theta_0^2}{cR_0}\bigg)\bigg(\cos l + \frac{\kappa^2}{\sin^2l + \kappa^2}\bigg), \\
    \label{eq:pbdot_z}
    (\dot{P}_{\rm b})_{\rm z} &= -1.08\times10^{-19}\frac{P_{\rm b}}{c}\bigg(\frac{1.25z}{[z^2 + 0.0324]^{1/2}} + 0.58z\bigg)\sin b, \\
    \label{eq:pbdot_mu}
    (\dot{P}_{\rm b})_\mu &= \frac{\mu^2d}{c}P_{\rm b},
\end{align}

\noindent and where: $n_{\rm b} = 2\pi/P_{\rm b}$ is the orbital frequency; $m_{\rm tot} = m_{\rm p} + m_{\rm c}$; $l$ and $b$ are the Galactic longitude and latitude, respectively; $\Theta_0$ and $R_0$ are the Galactrocentric circular-speed and distance parameters for the Solar-System barycenter, respectively \citep[e.g.,][]{rmb+14}; $\kappa = (d/R_0)\cos b - \cos l$; $z = d\sin b$ is the projected vertical distance of the pulsar-binary system from the Galactic plane; and $\mu = \sqrt{\mu_\lambda^2 + \mu_\beta^2}$ is the magnitude of proper motion. In this work, we used $\Theta_0$ = 236.9(4.2) km s$^{-1}$ and $R_0$ = 8.178(26)\footnote{The uncertainty we report for $R_0$ is the composite value determined after adding the statistical and systematic uncertainties reported by \citet{aab+19} in quadrature.} kpc as determined by \citet{aab+19}.

In Equation \ref{eq:pbdot_z} we chose a model for quantifying the Galactic gravitational potential that is traditionally used in pulsar-timing studies \citep{kg89}. While other models have been developed and studied in recent pulsar literature, the $\sim$20\% uncertainty in (\pbdot)$_{\rm obs}$ is too large for resolving statistically meaningful differences between model predictions \citep{pb18}. 

While expected to be negligible, we included the (\pbdot)$_{\rm GR}$ term in our analysis for completeness, as it has been observed in other pulsar-binary systems. \citet{hkw+20} presented equations for additional sources of \pbdot{} corrections that can eventually be observed through pulsar timing, arising from mechanisms such as mass loss and higher-order corrections predicted by GR. However, these additional terms are at least five orders of magnitude smaller than the current uncertainty in \pbdot, and we therefore chose to ignore those terms in subsequent calculations.

\begin{figure}
    \centering
    \includegraphics[scale=0.075]{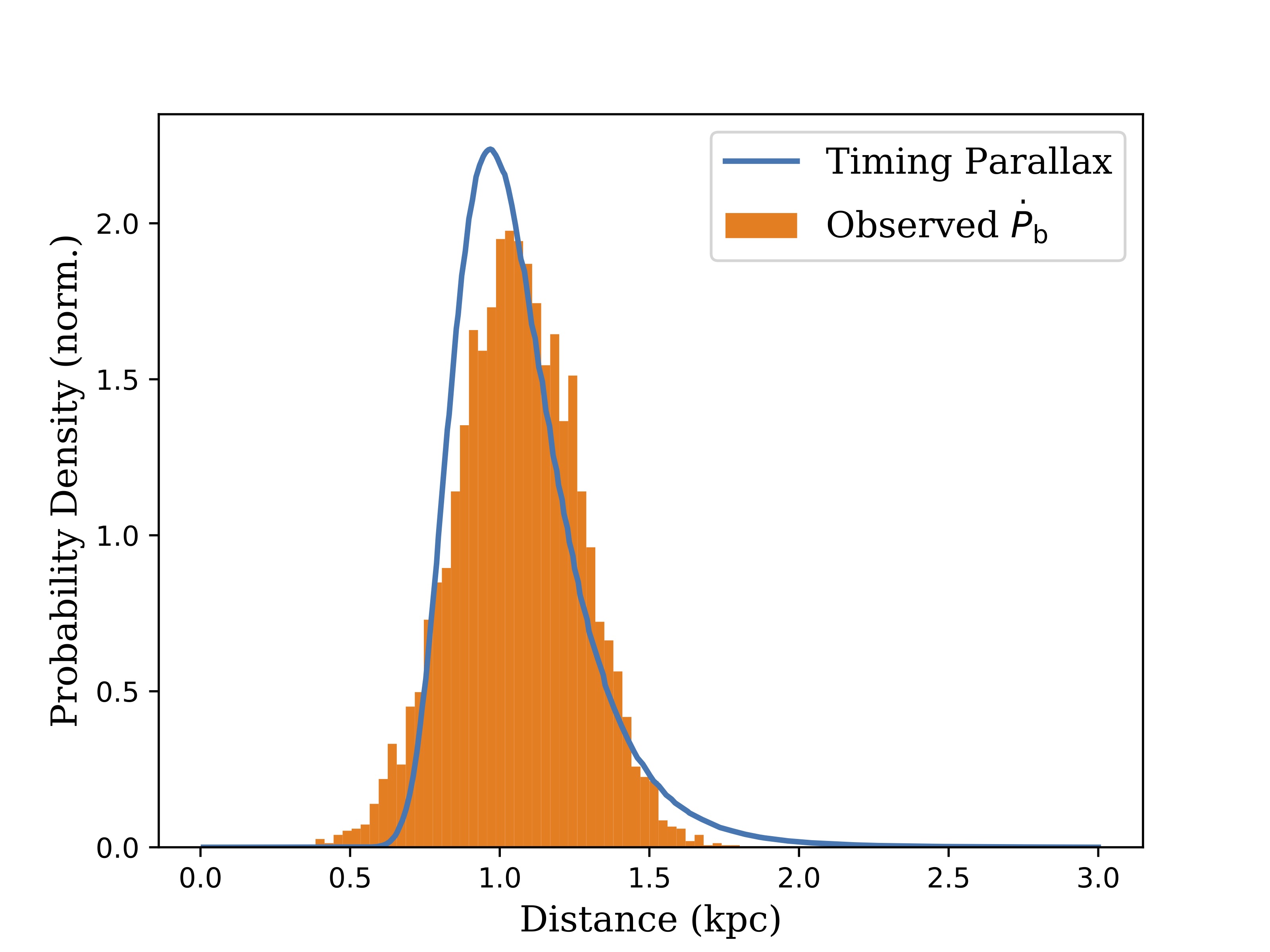}
    \caption{Independent estimates of distance from the extended timing data set for PSR J0740+6620. Shown above are distributions of distance to the PSR J0740+6620 binary system determined from the measured timing parallax (line) and Monte Carlo calculations of (\pbdot)$_{\rm obs}$.}
    \label{fig:distance}
\end{figure}

Equations \ref{eq:pbdot_GR}-\ref{eq:pbdot_mu} demonstrate that (\pbdot)$_{\rm obs}$ is ultimately a function of $m_{\rm p}$, $m_{\rm c}$, and $d$.\footnote{(\pbdot)$_{\rm obs}$ is also a function of several well-measured timing parameters, such as the orbital elements and astrometry. However, we held such values fixed to their best-fit estimates due to their high statistical significance.} To check for self-consistency in our best-fit timing models, we performed a Monte Carlo analysis of the relation \pbdottotal{} by sampling a set of values for \{(\pbdot)$_{\rm obs}$, $m_{\rm p}$, $m_{\rm c}$, $R_0$, $\Theta_0$, $\mu$\}, based on uncertainties obtained in this work or as reported in past literature, and solving for $d$ using a Newton-Raphson method. The distribution for $d$ obtained from a Monte Carlo simulation is shown in Figure \ref{fig:distance} along with the PDF for $d$ obtained from measurement of $\varpi = 1/d$. The overlapping PDFs demonstrate that the two independent timing effects yield statistically consistent estimates of $d$. The width of the PDF obtained from \pbdot{} is dominated by its measurement uncertainty, and width contributions from sampling the \{$R_0$, $\Theta_0$, $\mu$\} terms are negligible. We therefore held the \{$R_0$, $\Theta_0$, $\mu$\} terms fixed to their best-fit values in subsequent analysis.

The PDF derived from $\varpi$ was not corrected for the Lutz-Kelker bias in pulsar timing-parallax measurements, which arises in low-precision measurements of $\varpi$ and can lead to statistical underestimates of $d$ \citep{vlm10}. However, the consistency between the two distributions of $d$ for PSR J0740+6620 -- derived from independent timing effects -- indicates that the Lutz-Kelker bias is insignificant in our estimate of $\varpi$, and we therefore ignore this bias in subsequent analysis.

\subsection{Likelihood Analysis of the Shapiro Delay and System Distance}
\label{subsec:analysis_pdf}
For all timing models presented here, we used the $\chi^2$-grid method developed by \citet{sna+02} for determining posterior probability density functions (PDFs) and robust credible intervals for \{$m_{\rm p}$, $m_{\rm c}$, $i$, and $d$\}. The procedure we used for PDF and credible-interval estimation is described as follows:

\begin{itemize}
    \item selected a set of values for $m_{\rm c}, \cos i = \sqrt{1 - \sin^2i}$, and $d$, each from uniform distributions;
    \item computed values of $\sin i$, $m_{\rm p}$, $\varpi = 1/d$, and (\pbdot)$_{\rm tot}$ using Equations \ref{eq:massfunc}-\ref{eq:pbdot_mu}, based on the current location in the ($m_{\rm c}$, $\cos i$, $d$) phase space; 
    \item held the corresponding values of \{$\varpi$, $m_{\rm c}$, $\sin i$, (\pbdot)$_{\rm tot}$\} fixed in the timing model;
    \item re-fitted the timing model, allowing all timing-model parameters not defined on the grid to remain unconstrained during the fit;
    \item recorded the best-fit $\chi^2$ value, and repeated the above steps for different values of ($m_{\rm c}$, $\cos i$, $d$).
\end{itemize}

We ultimately obtained a three-dimensional grid of $\chi^2$ values computed over uniform steps in ($m_{\rm c}$, $\cos i$, $d$), which were then mapped to a likelihood function $p({\rm data, DMX}_j|m_{\rm c},\cos i, d) \propto \exp(-\Delta\chi^2/2)$ where $\Delta\chi^2 = \chi^2 - \textrm{ min}(\chi^2)$. The notation ``DMX$_j$" refers to one of the three DMX models, labeled with subscript $j$, that was generated for this work. We then used Bayes' theorem to compute the posterior distribution $p(m_{\rm c},\cos i, d|{\rm data, DMX}_j)$ for uniform priors on the physical parameters and choice of DMX model.

\begin{figure}
    \centering
    \includegraphics[scale=0.075]{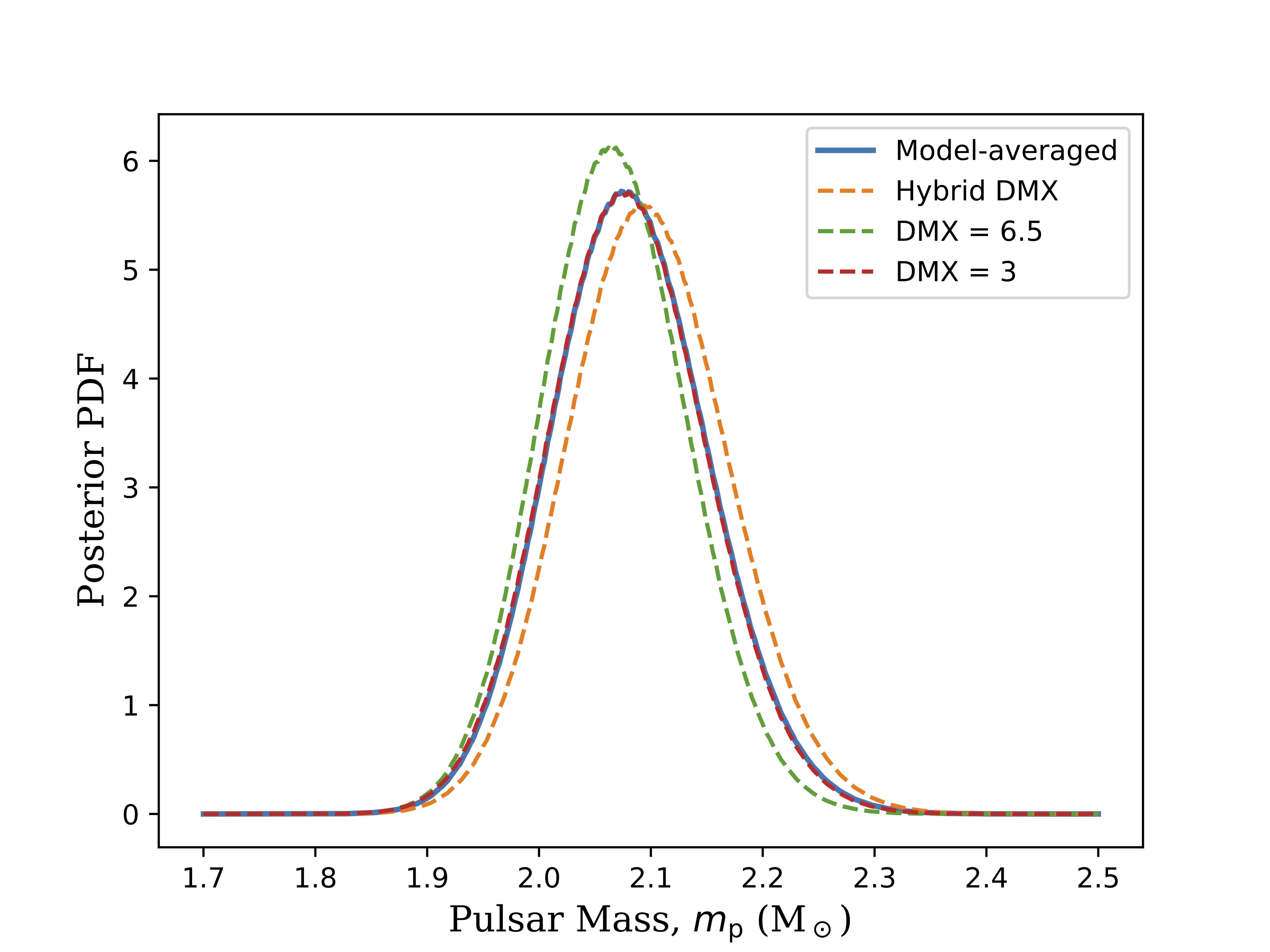}
    \caption{A comparison of normalized posterior PDFs of $m_{\rm p}$ estimated from the $\chi^2$-grid method applied to timing models of different DMX binning choices (dashed lines). For reference, the resultant posterior PDF obtained using the Bayesian model averaging method described in Section \ref{subsec:analysis_pdf} is shown as the red-solid line, and largely overlaps the orange-dashed curve.}
    \label{fig:compare_PDFs}
\end{figure}

Systematic uncertainties in the ($m_{\rm p}, m_{\rm c}, i, d$) parameters may arise due to choices in modeling DM variations. We used Bayesian model averaging \citep[e.g.,][]{hmrv99} to obtain PDFs and credible intervals that better reflect the model-independent distributions, in order to address DMX-model uncertainty. The Bayesian model averaging method defines the model-independent PDF as a weighted summation of model-dependent PDFs; in the case of the three-dimensional posterior distribution described above, the model-averaged PDF is 

\begin{align}
    \label{eq:bma}
    p(m_{\rm c},\cos i, d|{\rm data}) &= \sum_j p(m_{\rm c},\cos i, d|{\rm data, DMX}_j) \nonumber \\
    &\phantom{{}={}} \times p({\rm DMX}_j|{\rm data})
\end{align}

\noindent where $p({\rm DMX}_j|{\rm data})$ is the conditional posterior PDF for model DMX$_j$. In the absence of any preference in DMX modeling, all DMX models are equally likely and Equation \ref{eq:bma} reduces to a straightforward averaging of the three normalized PDFs. We used Equation \ref{eq:bma} to compute model-averaged posterior PDFs and their corresponding credible intervals for all four gridded parameters. The DMX$_j$ and model-averaged posterior PDFs for $m_{\rm p}$ are shown in Figure \ref{fig:compare_PDFs}.

Posterior PDFs and credible intervals for $m_{\rm p}$ were derived from the  original three-dimensional posterior PDF by noting that

\begin{align}
    \label{eq:translatedtoMP}
    p(m_{\rm p}, \cos i, d|{\rm data}) &= p(m_{\rm c}, \cos i, d|{\rm data}) \bigg|\frac{\partial m_{\rm c}}{\partial m_{\rm p}}\bigg|, \\
    \label{eq:partialmass}
    \textrm{where } \frac{\partial m_{\rm c}}{\partial m_{\rm p}} &= \frac{2f_mm_{\rm tot}}{3m_{\rm c}^2\sin^3i - 2f_mm_{\rm tot}},
\end{align}

\noindent and where Equation \ref{eq:partialmass} is determined by the Keplerian mass function ($f_m$). Using these original and translated PDFs, we computed credible intervals by marginalizing over the relevant subset of parameters, and also calculated two-dimensional posterior PDFs for all possible parameter pairs. We ignored the uncertainty in $f_m$ for all PDF calculations due to its negligible contribution to PDF widths \citep[relative uncertainty $\sim 10^{-7}$;][]{cfr+20}. Table \ref{tab:summary_pdfs} lists the credible intervals obtained using the methods described above.

%--------------------------------------------------------------------
\section{Discussion}
\label{sec:discussion}

The combined timing data set for PSR J0740+6620 contains new information on temporal and DM variations, as well as improved constraints of effects previously reported by \citet{cfr+20}. We discuss various aspects of these features in detail below, using data products generated with the techniques outlined in Section \ref{sec:analysis}. These data products are publicly available and include an initial set of posterior-PDF estimates used by \nicer{} to inform their modeling of surface thermal emission with our improved estimates for the PSR J0740+6620 system parameters \citep{zen21}. Our updated constraints on mass, geometric and distance parameters were used by \nicer{} to constrain the radius of PSR J0740+6620; their application is described elsewhere \citep{mld+21,rwr+21,raj+21}.

\subsection{Timing Properties in the \chimepsr{} Era}
All best-fit timing models developed for this work yielded consistent weighted root mean square (RMS) residuals: $\sim$0.3 \us{} for the GBT/1400-MHz data; $\sim$0.4 \us{} for GBT/820-MHz; and $\sim$1 \us{} for \chimepsr{} data. The factor of $\sim$2 difference in RMS residual between \chimepsr{} and GBT data is consistent with the expected timing precision based on observed sensitivity of the \chimepsr{} system; a recent timing analysis of PSR J0645$+$5158, another MSP observed with the GBT and \chimepsr{}, yielded similar RMS statistics \citep{abb+20}. 

The white-noise model of the combined TOA set, determined using the methods outlined in Section \ref{subsec:methods_noise}, is consistent with the model developed by \citet{cfr+20}. Moreover, the \chimepsr{} data set yields TOA-uncertainty scale and quadrature values comparable to those obtained for the NANOGrav data set. Therefore, while slightly less precise than its NANOGrav counterpart, the \chimepsr{} instrument is producing TOAs with noise properties that are consistent with behavior observed when using other observatories.

The GBT timing data currently possess greater statistical weight on parameter constraints due to their higher timing precision and larger number of TOAs. It is worth noting that a timing analysis of the \chimepsr{} data set on its own yields robust (albeit weaker) measurement of the Shapiro delay, with $m_{\rm c} = 0.28(2)$ M$_\odot$ and $\sin i = 0.9989(8)$ that correspond to $m_{\rm p} = 2.4(3)$ M$_\odot$. These CHIME/Pulsar estimates are consistent with the combined-set values obtained using the methods discussed in Section \ref{sec:analysis}. The high-cadence nature of \chimepsr{} observations, along with the 4.8-day orbit of PSR J0740+6620, has led to a statistically significant constraint on the Shapiro delay with only $\sim$1 year of timing data, considerably faster than the rate that was achieved with the GBT. Nonetheless, a combination of high cadence and high sensitivity is the only way to meaningfully improve the mass and geometric estimates of the PSR J0740+6620 system. We discuss these prospects further below.

\subsection{DM Variations Toward PSR J0740+6620}
As shown in Figure \ref{fig:residuals}, the DM of PSR J0740+6620 varies smoothly and slowly across the full data set, with few outlying points and no sign of annual variations due to interactions with free electrons of the Solar wind. These features are consistent with those reported by \citet{dvt+20}, who analyzed a 4.8-yr data set of PSR J0740+6620 acquired using the German LOng Wavelength (GLOW) consortium of telescope stations built for the LOw Frequency ARray (LOFAR), which were all sensitive around $\sim$150 MHz. The lack of periodic solar-wind variations is expected given the high ecliptic latitude of the PSR J0740+6620 system ($\beta \approx 44^\circ$, see Table \ref{tab:timing_model}), which leads to negligible traversal of the pulsar signal through the circumsolar medium.

The differences in frequency coverage and timing precision between the GBT and \chimepsr{} data sets lead to observable differences in their DM measurements. In particular, while the GBT data yield superior RMS timing residuals in both receiver bands than those obtained with \chimepsr, the \chimepsr{} data nonetheless yield better precision in DM measurements; the median uncertainty in GBT DMs determined with \pp{} is $\sim8\times10^{-4}$ pc cm$^{-3}$, while the same measure in the \chimepsr{} data set is $\sim2\times10^{-4}$ pc cm$^{-3}$. The median DMX measurement uncertainty in the \chimepsr{} era has comparable precision to that obtained with GLOW by \citet{dvt+20}, $\sim4\times10^{-5}$ pc cm$^{-3}$. Further analysis of the DM timeseries for PSR J0740+6620 will be the subject of future works.

\subsection{Updated Estimates of the Shapiro Delay Parameters}

\begin{deluxetable*}{l|c|c|c|c|c}
    \tablehead{\colhead{Parameter} & \colhead{\citet{cfr+20}\tablenotemark{a}} & \colhead{DMX=3.0} & \colhead{DMX=6.5} & \colhead{Hybrid-DMX} & \colhead{Model-Averaged}}
    \tablecaption{Credible intervals for the Shapiro-delay and distance parameters in the J0740+6620 system. \label{tab:summary_pdfs}}
    \startdata
    Pulsar mass, $m_{\rm p}$ (M$_\odot$) & $2.14^{+0.10}_{-0.09}$ & $2.08^{+0.07}_{-0.07}$ & $2.07^{+0.07}_{-0.06}$ & $2.10^{+0.07}_{-0.07}$ & $2.08^{+0.07}_{-0.07}$ \\
    Companion mass, $m_{\rm c}$ (M$_\odot$) & $0.260^{+0.008}_{-0.007}$ & $0.253^{+0.006}_{-0.005}$ & $0.252^{+0.005}_{-0.005}$ & $0.254^{+0.006}_{-0.006}$ & $0.253^{+0.006}_{-0.005}$ \\
    System inclination, $i$ (degrees) & $87.38^{+0.2}_{-0.2}$ & $87.53^{+0.17}_{-0.18}$ & $87.61^{+0.16}_{-0.16}$ & $87.53^{+0.17}_{-0.18}$ & $87.56^{+0.17}_{-0.18}$ \\
    Distance, $d$ (kpc) & $<4.4$\tablenotemark{b} & $1.16^{+0.17}_{-0.15}$ & $1.11^{+0.17}_{-0.15}$ & $1.15^{+0.17}_{-0.15}$ & $1.14^{+0.17}_{-0.15}$ \\
    \enddata
    \tablenotetext{a}{Values initially determined by \citet{cfr+20} are listed for comparison. The $\chi^2$ gridding procedure used by \citet{cfr+20} only considered one DMX model, with maximum bin size of 6.5 days.}
    \tablenotetext{b}{Neither \pbdot{} nor timing parallax -- and thus the distance -- were significantly constrained with the data set presented by \citet{cfr+20}. We instead quote here the $2\sigma$ upper limit obtained from their reported constraint.}
\end{deluxetable*}

\begin{figure*}
    \centering
    \includegraphics[scale=0.145]{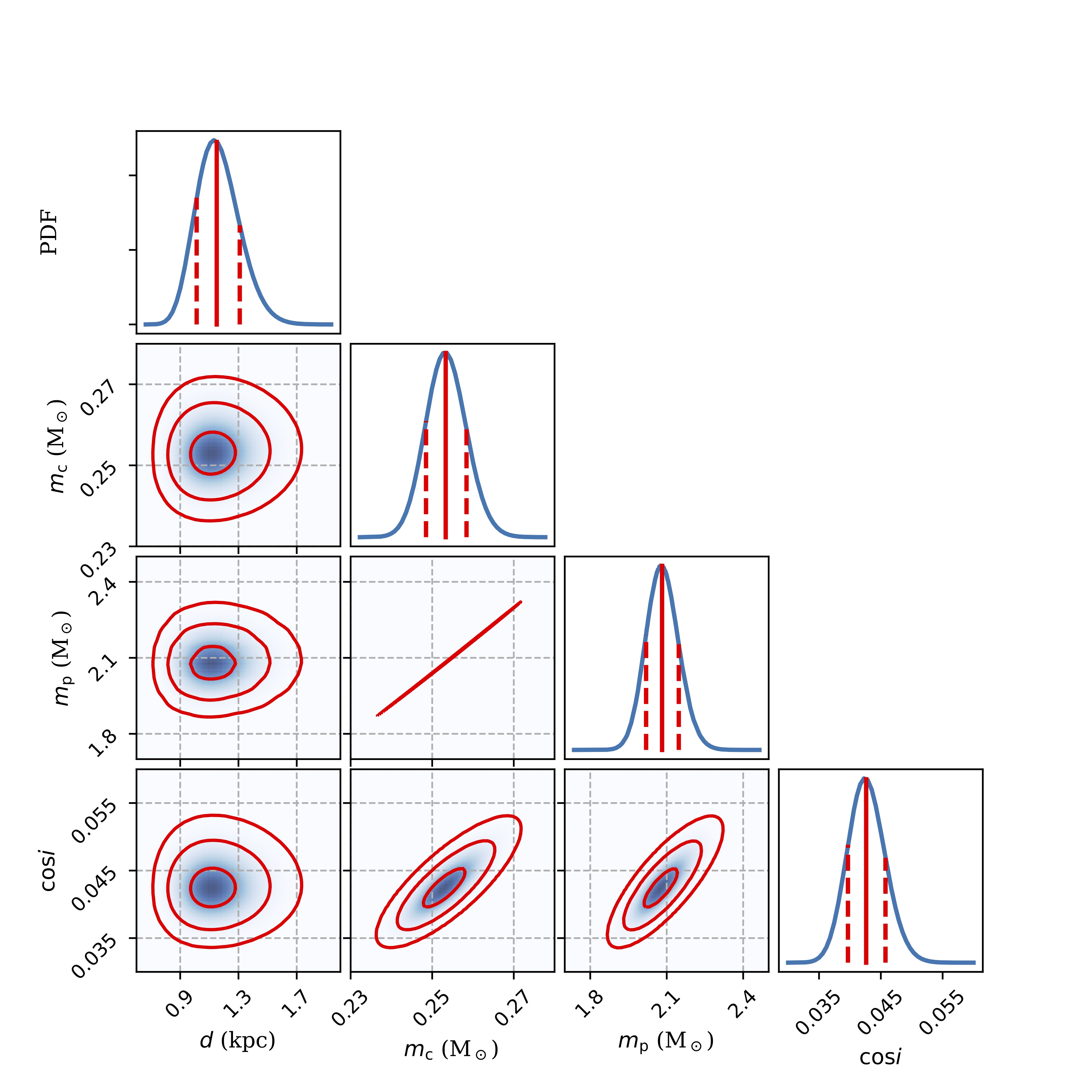}
    \caption{A triangle plot of two-dimensional, model-averaged posterior PDFs (off-diagonal) and marginalized, one-dimensional PDFs (on-diagonal) of parameters that characterize the Shapiro delay and distance to the PSR J0740+6620 system, derived using the methods outlined in Section \ref{subsec:analysis_pdf}. Vertical red lines in each diagonal panel represent the median value (solid) and 68.3\% credible interval (dashed). Inner- to outer-most contours denote the 68.3\%, 95.4\% and 99.7\% regions of credibility, respectively.}
    \label{fig:grid_triangle}
\end{figure*}

We performed the $\chi^2$-grid and PDF analyses outlined in Section \ref{subsec:analysis_pdf} for all three DMX timing solutions derived from the combined NANOGrav+\chimepsr{} data set for PSR J0740+6620. A summary of the credible intervals derived from all posterior PDFs is presented in Table \ref{tab:summary_pdfs}. The model-averaged PDFs and credible intervals are shown in Figure \ref{fig:grid_triangle}.

For all four parameters defined on each $\chi^2$ grid, the three posterior PDFs are consistent with each other at 68.3\% credibility as well as with the estimates made by \citet{cfr+20}. Moreover, the extent of each credible interval is largely unaffected by the choice of DMX bin width. This consistency indicates that the choice of DMX model with different bin widths makes no statistical difference to the significance of our estimates in the PSR J0740+6620 system. However, it is also likely that this consistency is specific to PSR J0740+6620 due to its slowly-varying DM evolution being adequately modeled with coarser piecewise-constant DMX models.

Our 68.3\% credible interval of the model-averaged $m_{\rm p} = 2.08^{+0.07}_{-0.07}\textrm{ M}_\odot$ is reduced by $\sim$30\% in comparison to the estimate made by \citet{cfr+20}. The model-averaged $m_{\rm p}$ remains the largest of all other precisely measured pulsar masses determined with the Shapiro delay to date. The 95.4\% lower bound on the model-averaged $m_{\rm p}$, 1.95 M$_\odot$, is similar to the 1.96 M$_\odot$ lower bound initially determined by \citet{cfr+20}. The lower limit on the maximum mass of neutron stars from the PSR J0740+6620 system therefore remains unchanged in our analysis.

The model-averaged estimate of the companion mass, $m_{\rm c} = 0.253^{+0.006}_{-0.005}\;{\rm M}_\odot$, remains largely consistent with the prediction from expected correlations between companion masses and orbital sizes that arise due to extended periods of mass transfer \citep{ts99}. The model-averaged credible interval on $m_{\rm c}$ is in improved agreement with correlation parameters that define the $m_{\rm c}-P_{\rm b}$ relation for low-metallicity progenitors (i.e., metallic mass fraction $Z \sim 10^{-3}$ or lower), which was noted by \citet{cfr+20} to yield an expected $m_{\rm c} \sim0.25\;{\rm M}_\odot$. \citet{enbd20} demonstrated through numerical calculations that only metal-poor progenitors with helium interiors can donate matter and evolve to yield the current masses of the PSR J0740+6620 system, and our improved measurement of $m_{\rm c }$ further supports this conclusion.

\subsection{Distance to the PSR J0740+6620 System}

We obtained a constrained, model-averaged distance of $d = 1.14^{+0.17}_{-0.15}$ kpc from the combined NANOGrav and \chimepsr{} TOA data set when using the $\chi^2$-grid method described in Section \ref{subsec:analysis_pdf}. This updated distance estimate is consistent with the distance of $d \approx 0.9$ kpc derived from the observed mean DM, placement within the Milky Way Galaxy, and the model of Galactic electron number density developed by \citet{ymw16}; the distance estimated by the number-density model of \citet{cl02}, $d \approx 0.6$ kpc, is less consistent, though underlying systematic uncertainties for both electron-density models correspond to $\sim 30$\% on $d$ and thus reduce the tension.

Our model-averaged estimate of $d$ is statistically consistent at the 2$\sigma$ level with the marginal estimate of $d$ derived from $\varpi = 0.5(3)$ mas made by \citet{cfr+20}. However, these two estimates are in tension with the first estimate of $d = 0.4^{+0.2}_{-0.1}$ kpc made from initial NANOGrav timing of PSR J0740+6620 \citep[derived from  $\varpi = 2.3(6)$;][]{abb+18}. An important distinction in these estimates is that the timing solution presented by \citet{abb+18} yielded only a marginal detection of the Shapiro delay parameters. \citet{abb+18} also employed an ``approximate" orthometric parameterization of the Shapiro delay to model the relativistic effect as a Fourier expansion about the orbital period using a finite number of harmonic terms. As noted by \citet{fw10}, prominent Shapiro delays from highly inclined binary systems -- as was first established for PSR J0740+6620 by \citet{cfr+20} -- are best modeled using the analytically exact expression predicted by general relativity, instead of a finite-term Fourier expansion that is more appropriate for low-inclination systems. The combination of a sparse, low-cadence data set and sub-optimal modeling likely led to an inaccurate distance estimate determined for PSR J0740+6620 by \citet{abb+18}.

\begin{figure}
    \includegraphics[scale=0.18]{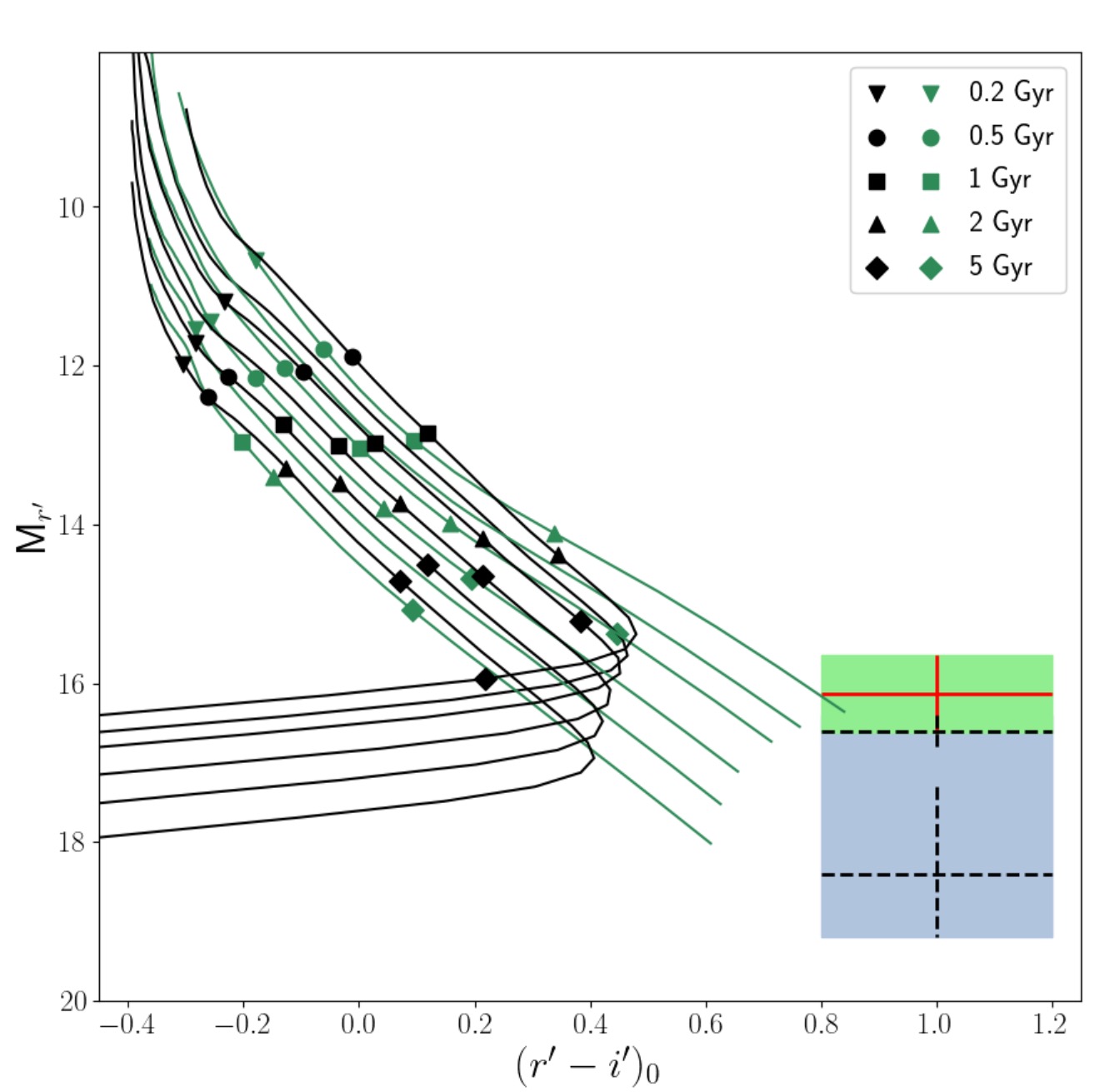}
    \caption{Color–magnitude diagram with white dwarf evolutionary sequences, evaluated at the $r'$ and $i'$ photometric bands. Black and green curves demonstrate the cooling predictions for white dwarfs with hydrogen and helium atmospheres, respectively \citep{hb06,ks06,tbg11,bwd+11}. For each colored set of curves, different tracks represent masses 0.2, 0.3, 0.4, 0.6, 0.8 and 1.0 M$_\odot$, with the mass increasing from upper to lower curves. Cooling ages along each track are marked by different symbols. The position of the companion to PSR J0740+6620 is indicated by the crosses: the dashed lower and middle crosses correspond to the latest DM distance estimate $d = 0.93$ kpc and the initial timing parallax distance $d = 0.4^{+0.2}_{-0.1}$ kpc from \citet{abb+18}, and the red cross demonstrates the companion location based on the new $d = 1.14^{+0.17}_{-0.15}$ kpc. The blue-shaded region shows the initial magnitude-distance uncertainty range as presented by \citet{bkk+19}, and the green-shaded region indicates the new uncertainty range corresponding to $d = 1.14^{+0.17}_{-0.15}$ kpc derived in this work. The two regions slightly overlap, and the green region covers the blue one.}
    \label{fig:mag}
\end{figure}

Finally, we reassessed the optical properties of the white-dwarf companion to PSR J0740+6620 using our direct measurement of $d$. \citet{bkk+19} concluded, based on their derived magnitudes and colors of the optical counterpart and the range of contemporaneous timing parallaxes and DM-based distances, that the pulsar companion is an ultra-cool, helium-atmosphere white dwarf with an effective temperature $<$ 3,500 K and cooling age $>$5 Gyr. However, \citet{bkk+19} were not able to constrain the mass of the companion due to the significant distance ambiguity. Observations with the {\it Gaia} satellite, designed for astrometric measurements, are currently not possible as the $r'$ and $i'$ apparent magnitudes of the white-dwarf companion to PSR J0740+6620 are $\sim26$ \citep{bkk+19}. These observed magnitudes are well above the limiting $g'$ magnitude of $\sim19$ for Gaia Data Release 2.\footnote{https://www.cosmos.esa.int/web/gaia/dr2}

Figure \ref{fig:mag} shows a comparison of the absolute magnitude and color of the optical source with cooling predictions for white dwarfs with hydrogen and helium atmospheres, and over a range of masses\footnote{http://www.astro.umontreal.ca/$\sim$bergeron/CoolingModels} \citep{hb06,ks06,tbg11,bwd+11}. The range of prior constraints on $d$ led to a large range in absolute magnitude, shown as the blue-shaded region in Figure \ref{fig:mag}, that encompasses cooling tracks of white dwarfs with $0.2-1\;{\rm M}_\odot$. Our updated measurement of $d$ restricts the absolute magnitude to the green-shaded region in Figure \ref{fig:mag} where only cooling curves for helium-atmosphere white dwarfs with masses $\sim 0.2-0.3 \;{\rm M}_\odot$ remain consistent with observations. This result is in agreement with the initial estimate of $m_{\rm c}$ made by \citet{cfr+20}, and with our updated $m_{\rm c}=0.253^{+0.006}_{-0.005} \textrm{ M}_\odot$. The improved synergy in optical- and radio-based estimates of intrinsic parameters for the PSR J0740+6620 system favors our new measurement of $d$, and further strengthens the counterpart association made by \citet{bkk+19}. 

Future optical and infrared studies with next-generation telescopes, combined with improved distance uncertainties from radio-timing observations, will be useful for further binary evolution modelling \citep{enbd20}. The increasing timespan of the PSR J0740+6620 data set will further improve the measurement of \pbdot{} and thus yield a precise kinematic distance, as was recently obtained for the PSR J1909$-$3744 binary system \citep{lgi+20}. Cooling predictions for white dwarfs with different composition, as well as stronger constraints on the distance and optical magnitudes, will allow for constraints on other properties of the white-dwarf companion, such as its hydrogen abundance. 

%--------------------------------------------------------------------
\section{Conclusions}
\label{sec:conclusions}

In this work, we extended the timing data set on the high-mass PSR J0740+6620 to incorporate additional GBT data acquired as part of the NANOGrav project, and to include the initial data set being generated by \chimepsr. We generated timing solutions for PSR J0740+6620 that were derived from both narrowband and wideband arrival times collected with the GBT and \chimepsr. For each data set, we also explored using three different piecewise-constant models of DM variations to assess sensitivity of the improved mass and geometric measurements on such modeling choices.

We found that all solutions were statistically consistent with one another at 68.3\% credibility, with minimal (i.e., $\sim$1\%) variation in credible intervals of the Shapiro delay parameters when comparing between several DM-evolution models. We obtained a model-averaged estimate of $m_{\rm p} = 2.08^{+0.07}_{-0.07}$ M$_\odot$, $m_{\rm c} = 0.253^{+0.006}_{-0.005}$ M$_\odot$ and $i = 87.56^{+0.17\circ}_{-0.18}$, consistent with the estimates made by \citet{cfr+20}. Our constraint on $m_{\rm p}$ is improved by $\sim$30\%, though the lower limit we derived for the maximum mass of neutron stars remains unchanged from that determined by \citet{cfr+20}. For the first time, our analysis also yielded a significant measurement of \pbdot{} which we argued is due to apparent acceleration from transverse motion. We found that the model-averaged $d = 1.14^{+0.17}_{-0.15}$ kpc when using the observed \pbdot{} as a constraint on the system distance, using the framework described in Section \ref{sec:analysis}.

We performed an independent timing analysis of TOA data recorded with the Nan\c{c}ay Radio Telescope (NRT) over a 7-yr timespan. All timing parameters presented in our work were statistically consistent with those estimated from NRT arrival times. The NRT data, along with forthcoming observations with the GBT and \chimepsr, will be the subject of future studies.

There are four ways in which our current estimates of the masses, geometry and distance can be significantly improved: continued, high-cadence timing with existing facilities; observations with ``ultra-wideband" radio receivers, which will become possible at the GBT in the next few years; observations with forthcoming telescopes of greater sensitivity and broadband coverage; and measurement of additional orbital variations that can be related to one or more of the fundamental parameters considered in this work. The most promising avenue for meaningful improvement on current constraints is the use of planned next-generation facilities. Several future radio observatories -- such as the Canadian Hydrogen Observatory and Radio-transient Detector \citep[CHORD;][]{vlg+19}, 2000-dish Deep Synoptic Array \citep[DSA-2000;][]{hrw+19}, and the next-generation Very Large Array \citep[ngVLA; e.g.,][]{cha18} -- will provide key opportunities in pulsar science with their heightened sensitivity and increased receiver bandwidths. 

For pulsar timing, the radiometer equation dictates the TOA uncertainty scales as $\sigma_{\rm TOA} \propto S_{\rm sys}/\sqrt{t_{\rm obs}\Delta f}$, where: $S_{\rm sys}$ is the system-equivalent flux density of the observatory; $t_{\rm obs}$ is the observation timespan, and $\Delta f$ is the receiver bandwidth.\footnote{For this calculation, we assume that intrinsic variations of pulse width and brightness across frequency are negligible. Our analysis of the combined NANOGrav and \chimepsr{} data sets, which collectively span the projected bandwidth of CHORD, largely support this assumption.} The expected improvement of TOA precision between \chimepsr{} \citep[with $S_{\rm sys} \approx 50$ Jy;][]{abb+20} and CHORD observations can be estimated based on current projections of CHORD design specification, with $S_{\rm sys} = 9$ Jy and $\Delta f =$ 1200 MHz \citep{vlg+19}. For observations of the same $t_{\rm obs}$, the band/time-averaged TOA uncertainty obtained with CHORD is $\sigma_{\rm TOA} \sim 0.1$ $\mu$s when using \chimepsr{} values listed in Table \ref{tab:summary_raw_data}. The DSA-2000 and ngVLA have similar planned bandwidths, but a factor of $\sim$ 5 smaller $S_{\rm sys}$ than that of CHORD. With CHORD projected to be operational by 2025, it is likely that daily TOA uncertainties of 0.1 $\mu$s and lower will be regularly achieved with next-generation observatories as soon as several years from now.

A reduction in $\sigma_{\rm TOA}$ by a factor of 10 or greater will lead to a similarly large improvement in the parameters of the Shapiro time delay. We therefore expect the credible interval of $m_{\rm p}$ to exceed values no larger than 0.01 M$_\odot$ of the median value with sufficient observations from CHORD, the DSA-2000 and/or ngVLA. The significance of \pbdot{}, and thus the constraint on distance, will continue to improve such that its relative uncertainty scales as $T^{-5/2}$, where $T$ is the data timespan \citep{dt92}. With ongoing observations and future prospects, we expect to gain additional, important insight from the PSR J0740+6620 system in the near future.

%--------------------------------------------------------------------
% add acknowledgements here.
\begin{acknowledgments}    
We acknowledge that CHIME is located on the traditional, ancestral, and unceded territory of the Syilx/Okanagan people. We thank Anne M. Archibald for useful discussion. We also thank the anonymous referee for their careful review of our work.

The NANOGrav project receives support from the National Science Foundation (NSF) Physics Frontiers Center (PFC) award number 1430284. The GBT is a facility of the NSF operated under cooperative agreement by Associated Universities, Inc. The National Radio Astronomy Observatory is a facility of the NSF operated under cooperative agreement by Associated Universities, Inc. We thank the telescope operators and all the staff at the GBT for the essential role they played in collecting data that were used in this work. 

We are grateful to the staff of the Dominion Radio Astrophysical Observatory, which is operated by the National Research Council of Canada.  CHIME is funded by a grant from the Canada Foundation for Innovation (CFI) 2012 Leading Edge Fund (Project 31170) and by contributions from the provinces of British Columbia, Qu\'ebec and Ontario. The CHIME Fast Radio Burst Project, which enabled development in common with the CHIME/Pulsar instrument, is funded by a grant from the CFI 2015 Innovation Fund (Project 33213) and by contributions from the provinces of British Columbia and Qu\'ebec, and by the Dunlap Institute for Astronomy and Astrophysics at the University of Toronto. Additional support is provided by the Canadian Institute for Advanced Research (CIFAR), McGill University and the McGill Space Institute thanks to the Trottier Family Foundation, and the University of British Columbia. The CHIME/Pulsar instrument hardware is funded by the Natural Sciences and Engineering Research Council (NSERC) Research Tools and Instruments (RTI-1) grant EQPEQ 458893-2014.

Support for HTC is provided by NASA through the NASA Hubble Fellowship Program grant \#HST-HF2-51453.001 awarded by the Space Telescope Science Institute, which is operated by the Association of Universities for Research in Astronomy, Inc., for NASA, under contract NAS5-26555. TTP is a NANOGrav PFC Postdoctoral Fellow and acknowledges support from the MTA-ELTE Extragalactic Astrophysics Research Group, funded by the Hungarian Academy of Sciences (Magyar Tudom\'anyos Akad\'emia), that was used during the development of this research. Portions of this work performed at NRL is supported by ONR 6.1 basic research funding. SMR is a CIFAR Fellow. Pulsar research at UBC is supported by an NSERC Discovery Grant and by CIFAR. TD and MTL are supported by an NSF Astronomy and Astrophysics Grant (AAG) award number 2009468. The Nan\c{c}ay Radio Observatory is operated by the Paris Observatory, associated with the French Centre National de la Recherche Scientifique (CNRS). We acknowledge financial support from the ``Programme National Gravitation, R\'ef\'erences, Astronomie, M\'etrologie'' (PNGRAM) of L'institut national des sciences de l'Univers (INSU) du CNRS, France. ECF is supported by NASA under award number 80GSFC17M0002. WF is supported by the STEM Mountains of Excellence graduate fellowship. DCG is supported by the John I. Watters research fellowship. VMK holds the Lorne Trottier Chair in Astrophysics \& Cosmology, a Distinguished James McGill Professorship and receives support from an NSERC Discovery Grant (RGPIN 228738-13) and Gerhard Herzberg Award, from an R.~Howard Webster Foundation Fellowship from CIFAR, and from the Fonds de Recherche du Qu\'ebec: Nature et technologies, Centre de Recherche en Astrophysique du Qu\'ebec. JWM is a CITA Postdoctoral Fellow, and this work was supported by NSERC (funding reference \#CITA 490888-16). 
\end{acknowledgments}

% cite software here.
\software{
    matplotlib \citep{hun07}, 
    \nanopipe{} \citep{dem18},
    \psrchive{} \citep{hvm04},
    \pp{} \citep{PP},
    \tempo{} \citep{nds+15},
}

% add biblio here.
\bibliographystyle{aasjournal}
\renewcommand{\bibname}{References}

\bibliography{main}

\end{document}